\documentclass[preprint,sort&compress,12pt]{elsarticle}

\usepackage{amssymb}
\usepackage{amsthm}
\usepackage{amsmath}
\usepackage{mathtools}
\usepackage{mathrsfs}
\usepackage{algorithm}
\usepackage[algo2e]{algorithm2e}
\usepackage{algpseudocode}
\usepackage{array}
\usepackage{multirow}
\usepackage{listings}
\usepackage{tabu}
\usepackage{booktabs}
\usepackage{enumerate}
\usepackage{fullpage}
\usepackage{float}
\usepackage{xcolor}
\usepackage[colorinlistoftodos]{todonotes}
\usepackage{bbm}
\usepackage{bm}
\usepackage[colorlinks=true]{hyperref}
\usepackage{url}
\usepackage{textcomp}
\usepackage{gensymb}
\usepackage{soul}
\usepackage{lineno}
\usepackage{graphicx}
\usepackage{subfigure}
\usepackage{float}
\usepackage{caption}
\usepackage{tikz}
\usetikzlibrary{shapes}

\theoremstyle{definition}

\theoremstyle{remark}

\biboptions{numbers,comma,round,square}
\graphicspath{ {./figs/} }

\journal{Elsevier}

\begin{document}
\begin{frontmatter}


\title{Diff-FlowFSI: A GPU-Optimized Differentiable CFD Platform for High-Fidelity Turbulence and FSI Simulations}




\author[cornellMAE,ndAME]{Xiantao Fan}
\author[ndAME]{Xinyang Liu}
\author[ndAME]{Meng Wang}
\author[cornellMAE,ndAME]{Jian-Xun Wang\corref{corxh}}

\address[cornellMAE]{Sibley School of Mechanical and Aerospace Engineering, Cornell University, Ithaca, NY, USA}
\address[ndAME]{Department of Aerospace and Mechanical Engineering, University of Notre Dame, Notre Dame, IN}

\cortext[corxh]{Corresponding author. Tel: +1 540 3156512}
\ead{jw2837@cornell.edu}
		
\begin{abstract}
Turbulent flows and fluid–structure interactions (FSI) are ubiquitous in scientific and engineering applications, but their accurate and efficient simulation remains a major challenge due to strong nonlinearities, multiscale interactions, and high computational demands. Traditional CFD solvers, though effective, struggle with scalability and adaptability for tasks such as inverse modeling, optimization, and data assimilation. Recent advances in machine learning (ML) have inspired hybrid modeling approaches that integrate neural networks with physics-based solvers to enhance generalization and capture unresolved dynamics. However, realizing this integration requires solvers that are not only physically accurate but also differentiable and GPU-efficient.
In this work, we introduce Diff-FlowFSI, a GPU-accelerated, fully differentiable CFD platform designed for high-fidelity turbulence and FSI simulations. Implemented in JAX, Diff-FlowFSI features a vectorized finite volume solver combined with the immersed boundary method to handle complex geometries and fluid–structure coupling. The platform enables GPU-enabled fast forward simulations, supports automatic differentiation for gradient-based inverse problems, and integrates seamlessly with deep learning components for hybrid neural–CFD modeling. We validate Diff-FlowFSI across a series of benchmark turbulence and FSI problems, demonstrating its capability to accelerate scientific computing at the intersection of physics and machine learning.

\end{abstract}

\end{frontmatter}


\section{Introduction} \label{s:intro}
Predictive modeling and simulations of complex fluid dynamics and their interactions with structures are fundamental in many engineering and scientific applications, including aerospace design, biomedical systems, climate modeling, and energy production. These physical processes are primarily governed by partial differential equations (PDEs), such as the Navier-Stokes equations for fluid dynamics and structural mechanics formulations for solid dynamics. Traditionally, these governing PDEs have been numerically solved using discretization methods such as finite difference, finite volume, or finite element techniques, collectively termed computational fluid dynamics (CFD). Over recent decades, CFD has evolved significantly, driven by advances in turbulence modeling, high-order numerical schemes, and sophisticated algorithms optimized for high-performance computing (HPC)~\cite{oberkampf2002verification,bhatti2020recent,zawawi2018review}.

Despite these advances, conventional physics-based CFD modeling exhibits fundamental limitations when applied to realistic, complex systems. Firstly, the underlying physics in many practical scenarios remain incompletely understood or cannot be fully resolved due to computational constraints. For instance, accurate simulation of multiphysics processes such as reactive flows, multiphase interfaces, or fluid–structure interactions (FSI) is still challenging, largely due to inadequate constitutive relations and closure models~\cite{ma2015using,fairbanks2020bi,sharma2024amrex,fuhg2024review, fan2021impacts}. Even canonical turbulent flows lack universally valid closure models, introducing significant model-form uncertainties~\cite{xiao2016quantifying,duraisamy2019turbulence,jofre2022rapid}. Secondly, CFD simulations typically rely heavily on precise knowledge of input parameters, boundary conditions, and initial conditions. However, these inputs are often uncertain or inaccessible in practical applications, significantly compromising predictive reliability and robustness. Lastly, the computational expense required for high-fidelity simulations presents a significant bottleneck. Resolving small-scale turbulence or complex multiphysics phenomena necessitates very fine spatial discretization and small integration time steps, substantially increasing computational cost. This computational burden severely restricts their practical applicability, particularly in iterative tasks such as optimization, control, or uncertainty quantification, where numerous repeated simulations are necessary.

In recent years, machine learning (ML), especially deep neural networks (DNNs), has emerged as a promising avenue to address these limitations by developing surrogate models that efficiently approximate complex mappings from data~\cite{brunton2020machine,du2022deep,han2022predicting,du2024confild}. However, purely data-driven ML approaches suffer from significant practical challenges, such as limited generalizability beyond their training regimes, dependence on extensive high-quality datasets, and a lack of physical interpretability. To mtigate these issues, incorporating physics to ML models emerges as a strategic solution, which is known as physics-informed machine learning (PIML)~\cite{karniadakis2021physics}. One of the notable PIML examples is physics-informed neural networks (PINNs)~\cite{raissi2019physics}, which incorporates physical laws into the training process through PDE-based residual losses. Although PINNs have shown considerable promise~\cite{sun2020surrogate,arzani2021uncovering,chen2020physics,gao2021phygeonet,li2025physics,chen2021physics}, the effectiveness of PINNs is often hinges upon meticulous hyperparameter tuning and suffers from convergence difficulties in training, particularly when simulating multipysics scenarios~\cite{krishnapriyan2021characterizing,wang2022and,wang2021understanding}. 

A promising alternative within the broader scope of PIML is the integration of ML directly with physics-based numerical solvers, forming hybrid neural–physics models. In these hybrid schemes, known physics prior are explicitly represented by discretized PDE operators, while DNN components typically learn unresolved or unknown dynamics within the PDEs systems. Early hybrid models typically adopted loosely coupled architectures, training ML components separately and embedding them subsequently into conventional numerical solvers~\cite{duraisamy2019turbulence,brenner2019perspective,wang2017physics,tompson2017accelerating}. However, these loosely coupled approaches have inherent limitations, notably compromised robustness, restricted adaptability, and limited generalizability, especially when dealing with long-term or strongly nonlinear multiphysics phenomena~\cite{wu2019reynolds,taghizadeh2020turbulence,fan2023differentiable}.

To address these issues, recent research has proposed the concept of \textit{neural differentiable modeling}, a framework that highlights the intrinsic connections between numerical PDE discretizations and common neural network architectures such as convolutional layers, graph kernels, and residual connections~\cite{liu2024multi,fan2023differentiable}. Within this perspective, conventional numerical solvers can be rigorously viewed as specialized neural architectures defined by established physics-based PDE operators~\cite{fan2025neural}. This unifying viewpoint facilitates a natural and strongly coupled integration between PDE-based numerical solvers and deep neural networks within a single, coherent differentiable programming ($\partial$P) framework. Differentiable programming, as a generalization of deep learning, supports end-to-end automatic differentiation (AD) through entire computational workflows, enabling holistic optimization and direct feedback between numerical PDE operators and neural network components. Such strongly coupled hybrid neural differentiable modeling frameworks significantly enhance the capability to learn from sparse or indirect observational data while retaining rigorous consistency with underlying physical laws. Recent applications of these neural differentiable models have demonstrated substantial improvements in predictive accuracy and modeling robustness across various scientific and engineering problems~\cite{kochkov2021machine,fan2023differentiable,fan2025neural,akhare2023diffhybrid,akhare2023physics,akhare2024probabilistic,list2022learned,belbute2020combining,shankar2025differentiable,shang2025jax}.

Central to hybrid neural differentiable modeling, differentiable CFD solvers provide an essential computational platform that seamlessly integrates numerical PDE operators with deep neural networks. Traditionally, gradient computations required for inverse modeling, optimization, or sensitivity analysis in CFD have been achieved through finite-difference approximations, adjoint methods, or source-to-source transformations~\cite{bischof2008implementation,mcnamara2004fluid,hinterberger2010automatic,shi2020natural,mader2008adjoint, mcnamara2004fluid}. However, these methods often face significant computational overhead, scalability and generalization issues, especially in large-scale or complex multiphysics simulations. Differentiable solvers developed within modern AD-enabled computational frameworks, such as JAX, PyTorch, or TensorFlow, allow for exact gradient computations without substantial additional computational cost. Recent developments, such as PhiFlow~\cite{holl2020phiflow}, JAX-CFD~\cite{kochkov2021machine}, and JAX-Fluids~\cite{RN1108}, exemplify the potential of differentiable CFD solvers to accelerate predictive modeling and enhance modeling robustness across various scientific and engineering tasks.

Nevertheless, the current state-of-the-art differentiable CFD frameworks remain limited primarily to simplified, single-physics scenarios, and their capabilities have not yet been rigorously extended to highly complex turbulent flows and multiphysics FSI problems. Particularly, simulations pose substantial challenges due to intrinsic nonlinearities, strong coupling effects, and multi-scale interactions between fluid and structural domains~\cite{boustani2021immersed}. Traditional CFD frameworks, often relying on the arbitrary Lagrangian–Eulerian (ALE) method~\cite{ramaswamy1987arbitrary}, face computational difficulties associated with frequent mesh regeneration and iterative solver updates, severely limiting their effectiveness in scenarios involving large structural deformations or highly dynamic interactions. Moreover, efforts to enhance existing CFD solvers using GPU acceleration, such as GPU-accelerated OpenFOAM variants, typically yield modest improvements (8\% speedup) due to partial GPU utilization and suboptimal integration with hardware architectures~\cite{malecha2011gpu,rathnayake2017openfoam}. These observations underscore the need for developing vectorized differentiable solvers from scratch, specifically designed to fully exploit GPU parallelism and memory hierarchies~\cite{harris2020array}.

Motivated by these critical limitations, we propose \textit{Diff-FlowFSI}, a GPU-optimized differentiable CFD platform explicitly designed for multiscale turbulent flow and strongly coupled FSI problems. Diff-FlowFSI harnesses the $\partial$P capabilities of JAX, providing an integrated platform to efficiently compute exact gradients and facilitate end-to-end optimization. It features a fully vectorized computational architecture designed explicitly for GPU parallelization, significantly enhancing computational efficiency and scalability. To circumvent computational challenges associated with mesh regeneration, Diff-FlowFSI employs the immersed boundary method (IBM), enabling efficient handling of large structural deformations and establishing a static computational graph that reduces runtime complexity~\cite{peskin2002immersed}. Furthermore, Diff-FlowFSI's differentiable nature inherently facilitates advanced inverse modeling, parameter identification, and optimization tasks by providing accurate gradient computations with respect to various simulation parameters. Crucially, Diff-FlowFSI serves as a versatile platform for constructing strongly coupled hybrid neural–physics models. By seamlessly embedding neural network components within its differentiable solver structure, Diff-FlowFSI can effectively address unresolved physical dynamics, poorly characterized boundary conditions, and model-form uncertainties directly from limited observational data. 

The main contributions of this paper include: (i) introducing the GPU-optimized fully-vectorized differentiable CFD solver Diff-FlowFSI, specifically designed for high-fidelity turbulent and multiphysics FSI simulations; (ii) demonstrating the solver's capabilities for efficient gradient computations, enabling robust inverse modeling and optimization tasks; and (iii) rigorously validating Diff-FlowFSI across multiple canonical turbulence and FSI benchmarks, highlighting substantial improvements in computational performance and predictive reliability compared to conventional methods.
The remainder of this paper is organized as follows. Section~\ref{s:meth} introduces the differentiable CFD methodology underpinning Diff-FlowFSI, including numerical details, GPU-oriented implementations, and automatic differentiation strategies. Section~\ref{s:validate} presents extensive validation results across canonical benchmark problems. Section~\ref{s:ad} demonstrates the application of Diff-FlowFSI to inverse modeling and hybrid neural differentiable modeling tasks. Finally, Section~\ref{s:conclusion} concludes the paper by summarizing key contributions and outlining directions for future research.

\section{Methodology} \label{s:meth}

\subsection{A differentiable and scalable CFD platform bridging physics and AI}

\begin{figure}[t!]
\centering
\includegraphics[width=\textwidth]{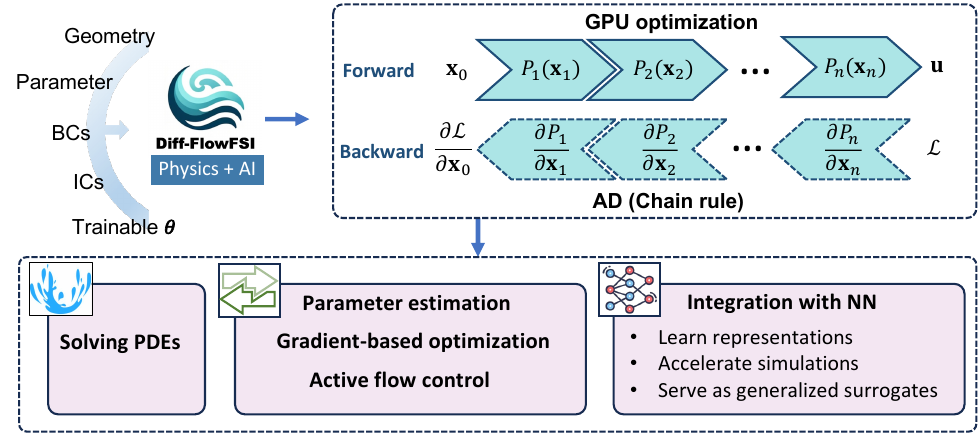}
\caption{Overview of the features and workflow of the differentiable CFD platform (Diff-FlowFSI).}
\label{fig:overview}
 \end{figure}

Diff-FlowFSI represents a significant advancement in CFD, leveraging the transformative potential of differentiable programming, vectorized numerical methods, and GPU acceleration (Figure~\ref{fig:overview}). At its foundation, Diff-FlowFSI is a high-fidelity, fully differentiable CFD solver capable of directly computing gradients through the entire simulation pipeline. This capability is transformative for optimization, sensitivity analysis, and inverse problems, enabling rigorous modeling of turbulence and FSI phenomena with high accuracy and efficiency. 

At the heart of Diff-FlowFSI is the formulation of the governing physics of fluid and structural dynamics, such as the Navier-Stokes equations and FSI coupling mechanisms, within the framework of differentiable programming ($\partial$P). $\partial$P generalizes deep learning (DL) by extending the power of neural networks to broader computational systems, enabling end-to-end gradient-based optimization for complex, physics-informed models. By leveraging AD, we can easily obtain the gradients of quantities of interest (QoIs) with respect to various inputs/parameters, such as boundary conditions, physical properties, or network trainable parameters. Unlike traditional CFD methods, which rely on external adjoint methods or heuristic optimizations, Diff-FlowFSI facilitates seamless gradient-based workflows while maintaining the physical fidelity inherent to classical numerical methods.

The platform’s vectorized architecture and GPU acceleration ensure computational efficiency, enabling scalable simulations of large-scale problems without compromising accuracy. By employing parallelized numerical kernels, optimized memory management, and vectorization, Diff-FlowFSI achieves high performance, even for computationally intensive problems such as high-resolution turbulence and complex FSI scenarios. This design ensures that the platform is not only robust but also practical for large-scale engineering applications.

A unique strength of Diff-FlowFSI lies in its extensibility. While it functions as a standalone differentiable CFD solver, it also supports seamless integration with modern AI techniques. This capability enables neural differentiable modeling~\cite{fan2023differentiable,fan2025neural}, where physics-based solvers are seamlessly combined with DL components to create hybrid models. Through $\partial$P, the hybrid neural modeling framework facilitates end-to-end training, allowing both the physical model and neural components to be optimized jointly, maintaining consistency and high robustness. This flexibility makes Diff-FlowFSI an ideal platform for tackling complex optimization and modeling challenges, particularly in scenarios where resolving full physics is infeasible, but sparse data are available to guide the modeling process.

The potential applications of Diff-FlowFSI span a wide range of fields, including aerodynamic design, flow control, biomimetic system optimization, and parameter inference. Its hybrid nature allows it to utilize both simulation and data to address forward and inverse problems with unprecedented efficiency and precision. By unifying the strengths of traditional numerical methods with the adaptability of modern AI, Diff-FlowFSI advances the state of CFD and establishes a versatile framework for bridging physics and machine learning. This platform exemplifies the promise of $\partial$P to revolutionize computational science. By marrying the rigor of physics-based solvers with the flexibility of deep learning, Diff-FlowFSI paves the way for next-generation computational tools that are both efficient and reliable. Next, we will explore the details of Diff-FlowFSI, including its governing equations, numerical methods, computation of gradients, GPU-optimized implementation within $\partial$P framework.  

\subsection{Governing equations} 
\label{s:sec2.1}

The incompressible Navier–Stokes (NS) equations governing fluid dynamics are given as,
\begin{equation}
\label{eq:ns}
\begin{aligned}
\nabla\cdot\mathbf{u} = 0, \hspace{7em}&\mathbf{x}, t \in \Omega_f \times [0, T]\\
\frac{\partial{\mathbf{u}}}{\partial t} = -(\mathbf{u}\cdot\nabla){\mathbf{u}}+\nu\nabla^2\mathbf{u}-\frac{1}{\rho}\nabla{p}+\mathbf{f}, \hspace{3em} &\mathbf{x}, t \in \Omega_f \times [0, T]
\end{aligned} 
\end{equation}
where $t$ and $\mathbf{x}$ denote time and spatial coordinates (Eularian), respectively; $\mathbf{u}(\mathbf{x},t)$ is the fluid velocity field, and $p(t, \mathbf{x})$ is the pressure field, both defined over the fluid domain $\Omega_f \subset \mathbb{R}^n$ with $n=2$ or $n=3$ for two- or three-dimensional problems, respectively. The constants $\rho$ and $\nu$ denote the fluid density and kinematic viscosity. The external body force $\mathbf{f}$ includes contributions from immersed structures, and the system is closed with appropriate initial and boundary conditions (ICs/BCs).

\begin{figure}[htp!]
\centering
\includegraphics[width=0.6\textwidth]{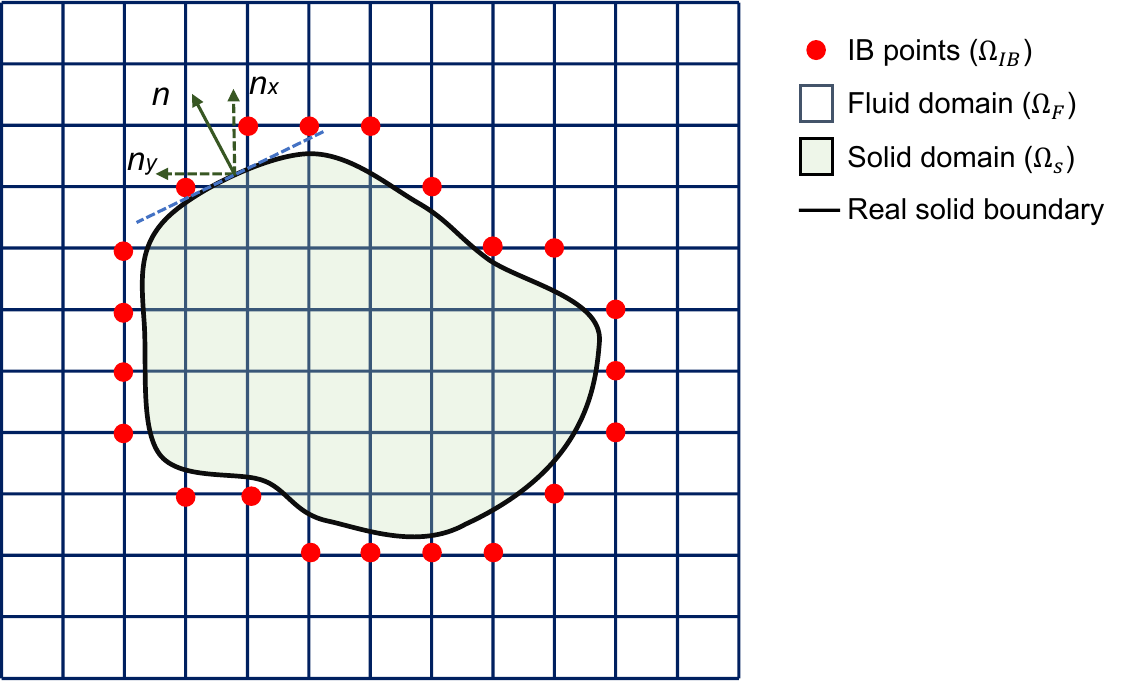}
\caption{Sketch of the immersed solid with sharp interface, where crosses represent cell centers.}
\label{fig:IBM}
 \end{figure}

The immersed solid is represented within a sharp-interface framework, as illustrated in Figure~\ref{fig:IBM}. The solid domain is defined as $\Omega_s$. Grid points in the fluid domain ($\Omega_f$) that have neighboring points inside $\Omega_s$  are classified as immersed boundary (IB) points ($\Omega_{IB}$); otherwise, they remain purely fluid points. The outward wall-normal vector at the interface is denoted by $\mathbf{n} = (n_x, n_y)^\top$. The geometry and motion of the immersed boundary are determined by the dynamics of the structure.  

The structural dynamics are governed in Lagrangian coordinates by
\begin{equation}
\label{eq: equation of solid}
\begin{aligned}
\mu_s \frac{\partial^2 \mathbf{w}}{\partial t^2}+EI \frac{\partial^4 \mathbf{w}}{\partial \mathbf{X}^4}=\mathbf{q}, \hspace{3em}&\mathbf{w}, t \in \Omega_s \times [0, T]
\end{aligned} 
\end{equation} 
where $\mathbf{X}$ is the Lagrangian coordinates; $\mathbf{w}(\mathbf{X},t)$ denotes the structural displacement defined within $\Omega_s \subset \mathbb{R}^2$; $\mu_s$ is the mass per unit length (or nondimensionalized mass-density ratio if scaled by fluid properties); $E$ is Young's modulus, $I$ is the second moment of area, and $EI$ is the bending stiffness (flexural rigidity). The term $\mathbf{q}(\mathbf{X}, t)$ represents the distributed forces induced by the fluid flow. Structures can be categorized as rigid and flexible; for two-dimensional problems, flexible structures are modeled as assemblies of rigid beam elements based on Euler-Bernoulli beam theory~\cite{RN1208}.

The coupling between the Eulerian (fluid) and Lagrangian (solid) descriptions is achieved through a discrete delta function, which facilitates interpolation and force spreading between the fluid and structure~\cite{peskin1972flow,uhlmann2005immersed}. This transformation is defined as, 
\begin{equation}
\label{eq:interpolation}
\begin{aligned}
    &\bm{\phi}(\mathbf{x},t) = \int_{\gamma_\mathbf{X}} \bm{\Phi}(\mathbf{X},t)\delta_h(\mathbf{x}-\mathbf{X})  d\gamma_\mathbf{X},  \hspace{3em} &\gamma_\mathbf{X} \in \Gamma \\
    &\bm{\Phi}(\mathbf{X},t) = \int_{\gamma_\mathbf{x}} \bm{\phi}(\mathbf{x},t)\delta_h(\mathbf{x}-\mathbf{X})  d\gamma_\mathbf{x} ,  \hspace{3em} &\gamma_\mathbf{x} \in \Gamma 
\end{aligned}
\end{equation}
where $\mathbf{\Phi}$ and $\mathbf{\phi}$ denote Lagrangian and Eulerian quantities, respectively; $\delta_h$ is the smoothed Dirac delta interpolation function; and the sets $\gamma_\mathbf{X}$ and $\gamma_\mathbf{x}$ denote the Lagrangian and Eulerian interfaces, respectively, along the immersed boundary $\Gamma$.

\subsection{Numerical algorithms} \label{s:sec2.2}

\subsubsection{Fluid discretization}

The fluid subsystem in Diff-FlowFSI is discretized using a finite volume method (FVM) on a staggered grid, where scalar quantities such as pressure are defined at cell centers and vector quantities such as velocity components are stored on their respective cell faces. This arrangement is particularly effective for satisfying the divergence-free condition in incompressible flows.  Both two-dimensional and three-dimensional configurations are supported, with options for uniform or non-uniform grid spacing.
Spatial discretization of the incompressible Navier–Stokes equations (Eq.~\ref{eq:ns}) is performed using second-order central differences for diffusive terms, and either a first-order upwind scheme or second-order central scheme for convective terms. For time advancement, we adopt explicit schemes such as forward Euler and Runge–Kutta methods, which are compatible with automatic differentiation.

To enforce incompressibility, we employ the Chorin projection method, a widely used fractional-step method. In this approach, the momentum equation is first advanced in time without enforcing the divergence-free constraint, yielding an intermediate velocity field $\mathbf{u}^*$:
\begin{equation}
\mathbf{u}^{*} = \mathbf{u}^t + \Delta t \left[ - \nabla \cdot (\mathbf{u}^{t} \otimes \mathbf{u}^{t}) + \nu \nabla^2 \mathbf{u}^{t} + \mathbf{f}^{t} \right],
\end{equation}
where $\Delta t$ is the time step, $\otimes$ denotes the tensor product operation, and $\mathbf{f}^t$ is the IBM forcing term introduced to enforce no-slip conditions on the immersed boundary.
The IBM is implemented using an adapted direct forcing strategy, where a penalization force is introduced only in cells intersecting the immersed region. This force ensures that the fluid velocity matches the solid velocity at the interface:
\begin{equation}
\mathbf{f}^{t}=\epsilon^t(\mathbf{x}) \left[\nabla\cdot({\mathbf{u}^t}\otimes{\mathbf{u}^t})-\nu\nabla^2\mathit{\mathbf{u}^t}+\frac{\mathbf{u}_s^{t}-\mathbf{u}^{t}}{\Delta t}\right],
\end{equation}
where $\epsilon^t(\mathbf{x})$ is the volume-of-solid function, with $\epsilon^t = 1$ in solid cells $\mathbf{x}_s \in \Omega_s$ and $0$ in fluid cells $\mathbf{x}_f \in \Omega_f$; $\mathbf{u}_s^t$ is the target velocity field of the fluid, introduced to facilitate the subsequent derivation. Specifically, $\mathbf{u}_s^t$ is interpolated from the structure velocity via Eq.~\ref{eq:interpolation} on the immersed boundary points ($\Omega_{IB}$), and is an artificial flow inside the solid region. This artificial flow is induced by the IBM and solved by the fluid equations.

To project the intermediate velocity onto a divergence-free space, we solve a Poisson equation for the pressure, assuming the corrected velocity satisfies,
\begin{equation}
\mathbf{u}^{t+1} = \mathbf{u}^{*} - \frac{\Delta t}{\rho} \nabla p^{t+1}.
\end{equation}
Taking the divergence yields the standard pressure Poisson equation:
\begin{equation}
\nabla^2p^{t+1}=\frac{\rho \nabla \cdot [\mathbf{u}^{*}- \mathbf{u}^{t+1}]}{\Delta t}.
\label{eq: initial possion}  
\end{equation}
The incompressibility condition $\nabla \cdot \mathbf{u}^{t+1} = 0$ holds in the fluid region $\Omega_f$, but does not apply in the solid domain $\Omega_s$, where the velocity field has no physical meaning. To maintain consistency across the fluid–structure interface, we introduce a surrogate divergence constraint in the solid region that penalizes deviations from the exact solid motion:
\begin{equation}
\label{eq:correction of incompressible}
\nabla \cdot \mathbf{u}^{t+1}= \nabla \cdot \left[ \epsilon^t(\mathbf{x})(\mathbf{u}_s^{t+1}-\mathbf{u}_c^{t+1}) \right] \approx \nabla \cdot \left[ \epsilon^t(\mathbf{x})(\mathbf{u}^{*}-\mathbf{u}_c^{t}) \right], 
\end{equation}
where $\mathit{\mathbf{u}_c^t}$ is the exact velocity of the immersed solid. This approximation modifies the divergence in the solid domain to reflect the discrepancy between the intermediate fluid velocity and the desired solid velocity. Substituting into the pressure projection yields a modified Poisson equation:
\begin{equation}
\nabla^2p^{t+1} = \frac{\rho\nabla \cdot \left[(1-\epsilon^t)\mathbf{u}^{*}+\epsilon^t \mathbf{u}_c^{t}\right]}{\Delta t}, \label{eq: pressure projection}   
\end{equation}
Note that this equation reduces to the classical pressure Poisson equation when $\epsilon^t = 0$ (i.e., in purely fluid regions). Finally, the corrected velocity field is updated as,
\begin{equation}
\label{eq: update velocity}
\mathbf{u}^{t+1}=\mathbf{u}^{*}  -\frac{\Delta t}{\rho}\nabla{p}^{t+1}.
\end{equation}

\subsubsection{Structure discretization}

The structural subsystem is discretized using the standard Galerkin finite element method~\cite{zienkiewicz2005finite}. The governing equation for structural dynamics (Eq.~\ref{eq: equation of solid}) is cast into a system of second-order ordinary differential equations using method of lines:
\begin{equation}
\mathbf{M} \frac{\partial^{2}\mathit{\mathbf{w}}}{\partial t^{2}}+ \mathbf{C} \frac{\partial \mathit{\mathbf{w}}}{\partial t}+\mathbf{Kw}=\mathbf{Q}. \label{eq: vibration equation}
\end{equation}
where $\mathbf{w} = [\mathbf{w}_1, \mathbf{w}_2, \cdots, \mathbf{w}_n]^T$ is the nodal displacement vector for $n$ Lagrangian structural nodes, and each $\mathbf{w}_i = [w_{x}^i, w_{y}^i, w_{\theta}^i]^T$ contains two translational components and one rotational degree of freedom. The matrices $\mathbf{M}$, $\mathbf{C}$, and $\mathbf{K}$ are the global mass, damping, and stiffness matrices, respectively. The fluid-induced forces $\mathbf{Q} = [\mathbf{q}_1, \mathbf{q}_2,\cdots,\mathbf{q}_n]^T$ are obtained through the spreading operation defined in Eq.~\ref{eq:fluid-to-solid}. As mentioned in Section~\ref{s:sec2.1}, two categories of structures are supported in Diff-FlowFSI: rigid and flexible bodies. The governing equations reduce to simplified oscillator models for rigid solids, while for flexible bodies, spatial discretization is carried out using beam elements, allowing for the resolution of structural deformation and modal dynamics. 

\begin{figure}[t!]
\centering
\includegraphics[width=0.9\textwidth]{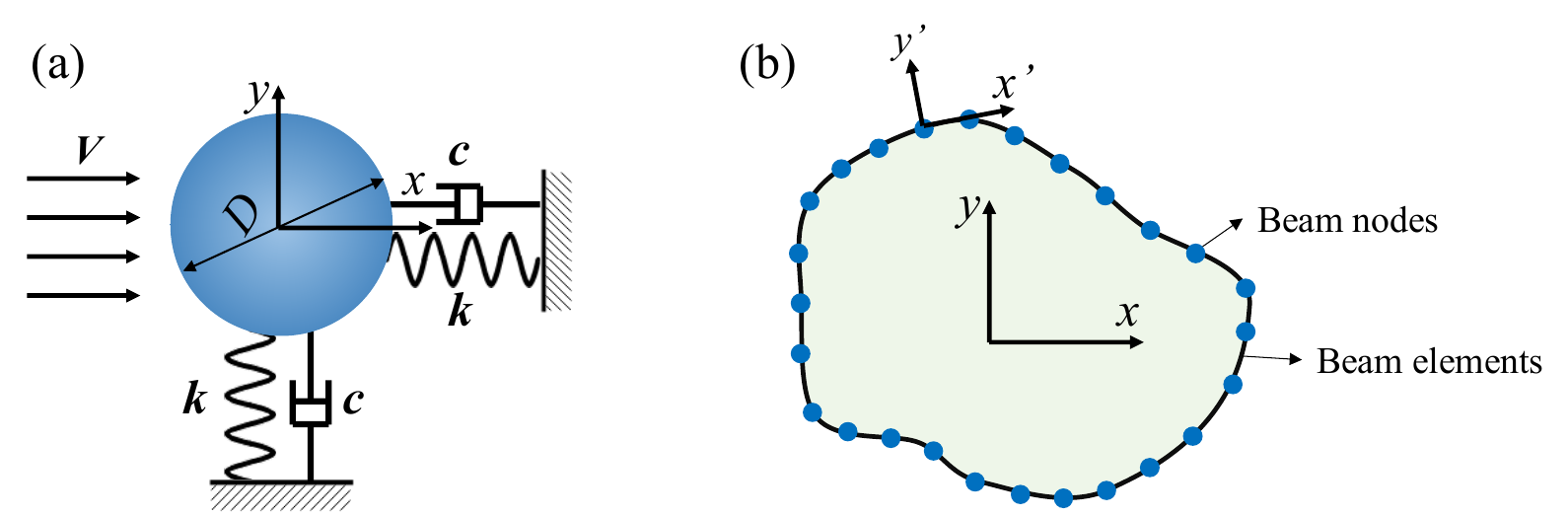}
\caption{(a) Mass-damper-spring system for rigid solids, allowing for an arbitrary solid shape. (b) Discretization of the solid into beam elements,, where $x-y$ is the global coordinate and $x'-y'$ is the local coordinate for each beam element.}
\label{fig:general_flexible}
 \end{figure}

\textit{\underline{Rigid solid:}} 
For rigid bodies, the structural dynamics reduce to a lumped mass-spring-damper oscillator model, as illustrated in Figure~\ref{fig:general_flexible}a. The system captures the translational response of the body subjected to hydrodynamic forces, such as drag and lift. This formulation is particularly suitable for modeling flow-induced vibrations (FIV), flapping hydrofoils, and aerodynamic oscillations of airfoils. In this case, the global mass, stiffness, and damping matrices $\mathbf{M}$, $\mathbf{K}$, and $\mathbf{C}$ in Eq.~\ref{eq: vibration equation} degenerate to scalar coefficients, leading to a simplified two-degree-of-freedom ODE system:
\begin{equation}
    m
    \left[
    \begin{array}{c}
         \ddot{w}_x^t  \\
         \ddot{w}_y^t 
    \end{array}
    \right]
    +
    k
        \left[
    \begin{array}{c}
         w_x^t  \\
         w_y^t 
    \end{array}
    \right]
    +
    c
        \left[
    \begin{array}{c}
         \dot{w}_x^t  \\
         \dot{w}_y^t 
    \end{array}
    \right]
    =
            \left[
    \begin{array}{c}
         F_D^t  \\
         F_L^t
    \end{array}
    \right].
\label{eq: rigid-body vibration}
\end{equation}
where $m$ is the effective mass; $k=2m \cdot (2\pi f_{n})^{2}$ is the stiffness with the the natural frequency $f_{n}$; $c=2m\cdot2\pi f_{n} \cdot \zeta_{s}$ is the damping coefficient with structural damping ratio $\zeta_{s}$; $F^t_D=\sum_{i=1}^{n} q^D_{i}$ and $F^t_L=\sum_{i=1}^{n}q^L_{i}$ denote the total drag and lift obtained by integrating $\mathbf{q}_i^t$ over all Lagrangian nodes, and $n$ is the total number of Lagrangian nodes. The resulting system is integrated using a fourth-order Runge–Kutta method.

\textit{\underline{Flexible solids.}} 
For flexible bodies, the domain is discretized into beam elements (Figure~\ref{fig:general_flexible}b), each governed by Euler–Bernoulli beam theory. In the 2D setting, only bending deformations are considered, and axial or shear effects are neglected. The global mass matrix $\mathbf{M}$ and stiffness matrix $\mathbf{K}$ in the Cartesian coordinate system $(x, y)$ are assembled from local element matrices $\mathbf{M}^e$ and $\mathbf{K}^e$ defined in the local beam coordinates $(x’, y’)$~\cite{RN1208}. The elemental stiffness matrix based on Euler–Bernoulli theory is given by:
\begin{equation}
 \mathbf{K}^e=\frac{EI}{L_e^3}
 \left[
 \begin{array}{c c c c}
 12 & 6 L_e & -12 & 6L_e \\
 6L_e & 4L_e^2 & -6 L_e & 2L_e^2 \\
 -12 & -6 L_e & 12 & -6 L_e \\
 6L_e & 2L_e^2 & -6 L_e & 4 L_e^2
 \end{array}
 \right],
 \label{eq:local stiffness}
\end{equation} 
where $L_e$ is the length of the element computed as $L_e=\sqrt{(x'_{i+1}-x'_i)^2+(y'_{i+1}-y'_i)^2}$. The corresponding consistent mass matrix for the beam element is:
\begin{equation}
 \mathbf{M}^e=\frac{\mu_s L_e}{420}
 \left[
 \begin{array}{c c c c}
 156 & 22 L_e & 54 & -13L_e \\
 22L_e & 4L_e^2 & 13 L_e & -3L_e^2 \\
 54 & 13 L_e & 156 & -22 L_e \\
 -13L_e & -3L_e^2 & -22 L_e & 4 L_e^2
 \end{array}
 \right],
 \label{eq:local mass}
\end{equation} 
 
Structural damping is modeled using Rayleigh damping~\cite{liu1995formulation}, where the global damping matrix is a linear combination of mass and stiffness:
\begin{equation}
\mathbf{C}=\alpha\mathbf{M}+\beta\mathbf{K},
\end{equation}
with damping coefficients $\alpha$ and $\beta$ defined from two selected modal frequencies $\omega_i$ and $\omega_j$ as,
\begin{equation}
    \alpha =2\zeta_s \dfrac{\omega_i \cdot \omega_j}{\omega_i + \omega_j}, \ \ \ \ \beta =\dfrac{2\zeta_s}{\omega_i + \omega_j},
\end{equation}
where $\zeta_s$ is the damping ratio. The modal frequencies are obtained by solving the generalized eigenvalue problem:
 \begin{equation}
     (\mathbf{K}-\bm{\omega}^2\mathbf{M})\mathbf{\Phi}=0.
 \end{equation}
where $\bm{\omega}=\mathrm{diag}[\omega_1 \dotsm \omega_i \dotsm \omega_j \dotsm \omega_n]$ is the the diagonal matrix of natural frequencies, and $\mathbf{\Phi}$ contains the corresponding vibration modes. The final system of second-order ODEs from Eq.~\ref{eq: vibration equation} is integrated in time using the Newmark–$\beta$ scheme, which balances numerical stability and second-order accuracy for structural dynamics.
 
\subsubsection{Two-way fluid-structure interactions} \label{s:sec2.3}

The coupling between the Eulerian fluid field and the Lagrangian structure is achieved through a two-way interpolation–spreading strategy using a smooth regularized Dirac delta function $\delta_h(\cdot)$. Specifically, following the continuous form in Eq.~\ref{eq:interpolation}, the velocity at each IB point is interpolated from the structural velocities using a half-distribution normalization to ensure sharp interface~\cite{ji2012novel},
\begin{equation}
\label{eq:solid-to-fluid}
    \mathbf{u}^t_{\mathrm{IB}}(\mathbf{x}) \cong \sum_{\mathbf{X}_i \in \gamma_{\mathbf{X}}} \frac{1}{F_{N_i}} \mathbf{\dot{w}}^t(\mathbf{X}_i) \delta_h(\mathbf{x}-\mathbf{X}_i) \Delta V(\mathbf{X}_i), 
\end{equation} 
where $\gamma_{\mathbf{X}}$ denotes the set of Lagrangian structural nodes near the IB point; $\Delta V(\mathbf{X}_i)$ is the effective boundary volume element associated with each Lagrangian point, approximated by a thin shell of thickness equal to one Eulerian grid spacing~\cite{uhlmann2005immersed}. The normalization factor $F_{N_i}$ is defined as
\begin{equation}
    F_{N_i}=\sum_{\mathbf{x} \in \gamma'_\mathbf{x}} \delta_h(\mathbf{x}-\mathbf{X}_i),
\end{equation}
where $\gamma'_{\mathbf{x}}$ is the set of surrounding Eulerian fluid nodes only outside the immersed boundary used in the interpolation.

The force exerted by the fluid $\mathbf{q}^t_i $ on the structure ($i^\mathrm{th}$ Lagrangian node) is obtained by spreading the direct forcing term $\mathbf{f}^t$ from the Eulerian grid to each Lagrangian node using the same kernel,
\begin{equation}
\label{eq:fluid-to-solid}
\mathbf{q}^t_{i}  \cong-\sum_{\mathbf{x} \in \gamma_\mathbf{x}} \mathbf{f}^t(\mathbf{x})   \delta_h(\mathbf{x}-\mathbf{X}_i) \Delta v(\mathbf{x}),
\end{equation}
where $\Delta v(\mathbf{x})$ is the control volume of the fluid cell centered at $\mathbf{x}$.

The regularized delta function is defined as a tensor product:
\begin{equation}
    \delta_h(\mathbf{x})= d_h(x)d_h(y),
\end{equation} \label{eq:smooth_function}
where $d_h(r)$ is a one-dimensional smoothing kernel:
\begin{equation}
\label{eq:delta_function}
    d_h(r)=\begin{cases} \frac{1}{8h}(3-2\frac{\rvert r \rvert}{h}+ \sqrt{1+4\frac{\rvert r \rvert}{h}-4(\frac{\rvert r \rvert}{h})^2}), & \rvert r \rvert \leq h\\\frac{1}{8h}(5-2\frac{\rvert r \rvert}{h}- \sqrt{-7+12\frac{\rvert r \rvert}{h}-4(\frac{\rvert r \rvert}{h})^2}), & h < \rvert r \rvert \leq 2h  \\0,&otherwise\end{cases}
\end{equation} 
where $h$ denotes the local fluid grid spacing.
Accordingly, the velocity field $\mathbf{u}(\mathbf{x}, t)$ across the entire computational domain is defined piecewise,
\begin{equation}
\label{eq:final velocity}
    \mathbf{u}(\mathbf{x}, t)=\begin{cases} \mathbf{u}(\mathbf{x},t), & \mathbf{x} \in \Omega_f  \\ \mathbf{u}_{\mathrm{IB}}, & \mathbf{x} \in \Omega_{IB} \\
    \overline{\mathbf{\dot{w}}}(\mathbf{X,t}), & \mathbf{x} \in \Omega_s
    \end{cases}
\end{equation}

To accurately capture dynamic interactions, especially for flexible structures with strong added-mass effects, a strongly coupled FSI scheme is implemented, as outlined in Algorithm~\ref{scheme: strong-coupling}. This algorithm iteratively solves the fluid and structure equations until convergence is achieved at each time step. The default convergence threshold is set to $\xi_t = 10^{-5}$ with a maximum of $N=100$ sub-iterations. Setting $N=1$ recovers a weakly coupled scheme.

\begin{algorithm}[t]
\footnotesize
\SetKwInput{KwInit}{Initialize}
\caption{Strong coupling procedure for Diff-FlowFSI solver} 
\For{$t \in [1,T]$} {
\KwInit{Initialize tolerance $\xi_t$ and step $N$}
$\xi_t \gets 1$ and $N \gets 1$ 

\While{$\xi_t \geqslant 1 \times 10^{-5}$ and $N \leqslant 100$}{
$\mathbf{u}^t \gets \mathcal{L} (\mathbf{u}^{t-1},\dot{\mathbf{w}}^{t-1}, p^{t-1}, \epsilon ^{t-1})$ \ \Comment{Update the fluid velocity based on Eq.~\ref{eq: pressure projection},~~\ref{eq: update velocity} and~\ref{eq:solid-to-fluid}}

$\mathbf{q}^t  \gets \mathcal{F} (\mathbf{f}^t(\mathbf{x})) $ \ \Comment{Calculate the fluid force based on Eq.~\ref{eq:fluid-to-solid}} 

$\mathbf{w}^t \gets \mathcal{G} (\mathbf{q}^t)$ \ \Comment{Calculate the structure responses based on Eq.~\ref{eq: vibration equation}}

$\xi_t \gets \parallel (\mathbf{w}^t -\mathbf{w}^{t-1})/\mathbf{w}^{t-1} \parallel^2_2$\ \Comment{Calculate the error}

$N \gets N+1$ \Comment{Update the sub-iteration step}

$\mathbf{w}^{t-1} = \mathbf{w}^t$ \ \Comment{Reset the structural responses}
}

}\label{scheme: strong-coupling}
\end{algorithm}

\subsection{Differentiable programming and GPU optimization}

Diff-FlowFSI is implemented entirely within the JAX framework to exploit the full potential of differentiable programming and GPU acceleration. This design allows not only high-performance forward simulations but also efficient, gradient-based optimization and learning for inverse problems, data assimilation, and hybrid neural-physics modeling. Here we describe the core algorithmic features that enable differentiability and GPU efficiency in Diff-FlowFSI, including AD modes, implicit differentiation, and GPU-oriented programming paradigms such as vectorization, JIT compilation, and scan-based loop unrolling.

\subsubsection{Automatic and implicit differentiation} \label{s:back_propa}

For a given mapping $f:\mathbb{R}^n \to \mathbb{R}^m$, the Jacobian matrix $\partial f \in \mathbb{R}^{m \times n}$ encodes the local sensitivity of outputs with respect to inputs, which is essential for gradient-based optimization for inverse problems and hybrid model training. JAX provides two primary AD primitives: Jacobian-vector products (JVPs) for forward-mode AD and vector-Jacobian products (VJPs) for reverse-mode AD. These enable efficient gradient computation depending on the relative dimensionality of the input and output spaces. In JVPs, the directional derivative $\partial f \cdot \mathbf{v}$ is computed for a known vector $\mathbf{v} \in \mathbb{R}^n$, producing an output in $\mathbb{R}^m$; whereas in VJPs, the adjoint derivative $\mathbf{v} \cdot \partial f$ is evaluated for $\mathbf{v} \in \mathbb{R}^m$, yielding gradients in $\mathbb{R}^n$.

Diff-FlowFSI primarily leverages reverse-mode AD via VJP to compute gradients of scalar loss functions $\mathcal{L}$ with respect to high-dimensional inputs, such as parameterized boundary conditions, physical parameters, or neural network weights. This setting, where $m=1$ and $n \gg m$, is common in data-driven optimization and physics-informed learning. In such cases, the full gradient $\frac{\partial \mathcal{L}}{\partial \bm{\theta}} = [\partial \mathcal{L}/\partial \theta_1, \dots, \partial \mathcal{L}/\partial \theta_n]$ is constructed column-wise through efficient reverse-mode backpropagation, avoiding explicit Jacobian storage and reducing memory cost~\cite{baydin2018automatic}. In contrast, forward-mode AD becomes advantageous when a small number of input variables influence a large number of outputs, such as when optimizing a low-dimensional design variable $\lambda \in \mathbb{R}^k$ ($k \ll m$) to control multiple physical objectives $\bm{E} \in \mathbb{R}^m$ (e.g., turbulent kinetic energy, structural stress, and vibration amplitude). JVPs allow computing $\frac{\partial \bm{E}}{\partial \lambda}$ one column at a time and are thus particularly effective for such low-input–high-output settings. For second-order optimization, Hessian-vector products can be obtained by nesting JVPs and VJPs, enabling curvature-aware updates without explicitly forming the Hessian.

Beyond explicit gradient paths, Diff-FlowFSI incorporates implicit differentiation~\cite{akhare2025implicit} to enable efficient and memory-optimal training of models involving fixed-point solvers or iterative subroutines. This is particularly critical for differentiating through inner solvers such as pressure Poisson equations, strongly coupled FSI iterations, or future extensions involving unrolled implicit time integration. Rather than tracking all intermediate steps of convergence, gradients are computed through the converged solution using the implicit function theorem~\cite{blondel2024elements-88d}. Specifically, for a fixed-point relation $\mathcal{F}(\textbf{u}, \bm{\theta})$, where $\mathbf{u}$ is the converged state and $\bm{\theta}$ are parameters, the total derivative of a loss function $\mathcal{L}(\mathbf{u}(\bm{\theta}))$ is given by,
\begin{equation}
    \frac{d \mathcal{L}}{d \boldsymbol{\theta}} = - \boldsymbol{\lambda}^\top \cdot \frac{\partial \mathcal{F}}{\partial \boldsymbol{\theta}}, \quad \text{where} \quad \left( \frac{\partial \mathcal{F}}{\partial \mathbf{u}} \right)^\top \boldsymbol{\lambda} = \left( \frac{\partial \mathcal{L}}{\partial \mathbf{u}} \right)^\top.
\end{equation}
This adjoint-based formulation circumvents backpropagation through the iterative loop, yielding exact gradients with minimal memory overhead. The linear system for $\bm{\lambda}$ can be solved using iterative methods and can exploit sparsity and structure in $\partial\mathcal{F}/\partial \mathbf{u}$. Diff-FlowFSI adopts this strategy via the JAXOPT library~\cite{jaxopt_implicit_diff}, which automates the application of the implicit function theorem in differentiable optimization routines. To further control memory usage in long unrolls, Diff-FlowFSI leveraged gradient checkpointing techniques, where only a subset of intermediate states are stored, and others are recomputed during backpropagation. This hybrid strategy of implicit differentiation and checkpointing enables robust training of hybrid models under severe memory constraints, especially in long-rollout, high-resolution spatiotemporal simulations.

\subsubsection{GPU-optimized implementation}\label{s:programm}

Conventional CFD solvers are architected for CPU-based execution, typically written in C/C++ or Fortran with optimization strategies tailored to serial or multi-threaded CPU pipelines. However, such CPU-oriented implementations are not suited for modern GPU accelerators, and they lack native support for automatic differentiation, limiting their use in inverse design, data assimilation, and integration with machine learning workflows. In contrast, Diff-FlowFSI is designed from the ground up to exploit the parallelism and memory bandwidth of GPU architectures. Implemented entirely in JAX, the framework enables composable kernel fusion, memory-efficient tensor operations, and seamless support for differentiable programming. This GPU-native design is essential not only for forward simulation performance but also for enabling scalable and memory-efficient training of hybrid neural-physics models. The following implementation strategies, including array-based vectorization, just-in-time (JIT) compilation, and scan-based loop unrolling, are central to achieving high throughput and memory efficiency on modern GPU accelerators.

\paragraph{Array programming for vectorization}
Unlike traditional loop-based implementations that rely on sequential execution, Diff-FlowFSI uses an array programming paradigm similar to NumPy~\cite{harris2020array}, in which operations are expressed over entire arrays or tensor batches. This eliminates the overhead of interpreted for-loops and allows full exploitation of GPU parallelism. Specifically:
\begin{itemize}
    \item \emph{Eulerian Grid Operations}: All spatial derivative evaluations and field updates are performed as element-wise or stencil-based operations over entire grids using vectorized computation.
    \item \emph{Eulerian–Lagrangian Coupling}: Interactions between Lagrangian markers and Eulerian fields, such as computing the volume-of-solid indicator ($\epsilon$) and regularized delta function ($\delta$) used in the immersed boundary force spreading, are implemented using \texttt{jax.vmap}, which vectorizes function application over batch dimensions for high-throughput execution.
\end{itemize}

\paragraph{Just-in-time (JIT) compilation} \label{s:jit}
To bridge the performance gap between Python and compiled code, Diff-FlowFSI makes extensive use of JAX's JIT compilation, which translates Python functions into highly optimized XLA kernels. This transformation enables not only significant speedups but also efficient memory usage by eliminating redundant allocations and enabling operation fusion. JIT compilation is especially beneficial for core solver subroutines that are repeatedly invoked over many time steps. In Diff-FlowFSI, the main simulation kernel is compiled once and reused across iterations, with input arguments structured as a Pytree to ensure static shape compliance, a requirement for compilation efficiency.


\paragraph{Scan-based loop unrolling}
To eliminate Python-side overhead and enable fully compiled rollout of iterative dynamics, Diff-FlowFSI replaces PDE-solving loops with JAX's \texttt{scan} primitive. Unlike a Python \texttt{for}-loop, \texttt{scan} promotes functional-style programming and enables loop body fusion and memory-efficient sequential execution within compiled code. More importantly, \texttt{scan} integrates seamlessly with JAX's AD and supports backpropagation through entire time horizons without excessive memory growth. Diff-FlowFSI employs a nested scan design:
\begin{itemize}
\item The inner scan (\texttt{fsi.funcutils.repeated}) evolves the system over a short burst of $N_1$ steps, storing only the final state.
\item The outer scan (\texttt{fsi.funcutils.trajectory}) chains $N_2$ such bursts and retains all intermediate outputs.
\end{itemize}
The overall memory cost thus scales with the grid size and outer loop depth $N_2$, but remains independent of $N_1$. To support longer simulations exceeding device memory, a third-level Python for-loop can be introduced, which invokes the outer scan repeatedly using the terminal state of each rollout as the initial condition for the next. This design is made modular and efficient through JAX's Pytree structure, which allows batched and structured updates of the solver state. The overall computational loop is summarized in Algorithm~\ref{al:main loop}.

\begin{algorithm}[t!]
    \footnotesize
    \SetKwInput{KwInit}{Initialize}
    \SetKwInput{KwSet}{Set}
    \SetKwInput{Kwfun}{One step function}
    \caption{Main computational loop in Diff-FlowFSI solver} 
    \KwInit{Wrap initial conditions as a Pytree $V_0$}
    \KwSet{Set inner steps $N_1$, out steps $N_2$ and for-loop iterations $N$}
    \Kwfun{Define the identical one-step function $f$ over the computational domain}
    \BlankLine    
    $\texttt{step\_fn}=\texttt{fsi.funcutils.repeated}(f, N_1)$\  \Comment{Wrap jax.scan function for the inner loop}
    \BlankLine    
    $\texttt{rollout\_fn} = \texttt{fsi.funcutils.trajectory(step\_fn}, N_2)$\  \Comment{Wrap jax.scan function for the outter loop}
    \BlankLine    
    $\texttt{rollout\_fn = jax.jit(rollout\_fn)}$\  \Comment{JIT compile}
    \BlankLine
    \For{$i \in [0, N]$} {
    \eIf{$i = 0$}{
    $V_0=V_0$      
    }
    {
    $V_0=\texttt{jax.tree\_util.tree\_map(lambda}~x: x[-1], \texttt{results})$ \ \Comment{Extract the pytree data}
    }
    $\texttt{results=rollout\_fn}(V_0)$ 
    }
    \label{al:main loop}
\end{algorithm}

\subsection{Supported Domains, Models, and Interfaces} \label{s:sec2.5}

Diff-FlowFSI is designed to support a wide range of computational settings for high-fidelity fluid and FSI simulations. Consistent with the numerical formulations described in previous sections, the solver operates on structured grids defined over rectangular domains. Both uniform and smoothly stretched grids are supported, enabling resolution adaptation near walls or immersed boundaries without modifying the static compute graph.

Boundary conditions are modular and extensible. As summarized in Table~\ref{tab: boundary}, the solver supports three fundamental types of boundary conditions, including periodic, Dirichlet, and Neumann, which can be flexibly applied to individual boundaries. These can be specified using constant values, spatially varying profiles, or time-dependent functions. Several canonical boundary configurations are pre-implemented, including periodic channel flow, no-slip and free-slip walls, velocity inlet with parabolic or sheared inflow, pressure outlet, convective outflow, and more. Synthetic inflow turbulence can be prescribed via procedural models.

Time integration is performed using either a fixed time step or an adaptive strategy governed by the Courant–Friedrichs–Lewy (CFL) condition based on the instantaneous maximum velocity. Both low-order (e.g., forward Euler) and high-order (e.g., explicit Runge–Kutta) schemes are available, allowing trade-offs between stability, accuracy, and cost.

For immersed boundaries, Diff-FlowFSI currently supports static solids in both two and three dimensions. Dynamic solids governed by Newtonian rigid-body or flexible structural dynamics are implemented in 2D, with coupling performed using strong two-way interactions. Extension to 3D dynamic solids is planned for future releases.

Turbulence modeling is built via optional subgrid and wall models. A constant Smagorinsky model is available for large-eddy simulation (LES), and an equilibrium wall shear stress model is provided for wall-modeled LES (WMLES) in high-Reynolds-number regimes.
\begin{table}[H]
\centering
\small
    \caption{Overview of numerical models, boundaries and technical features in Diff-FlowFSI}
    \begin{tabular}{lll}
    \toprule[1pt]
        Category & Type & Details \\ \midrule
        Grid & Uniform $\&$ Stretching & Static compute graph\\\midrule
        Time integration & Forward Euler & $1^{st}$-order Explicit scheme \\ 
        ~ & Runge Kutta & High order explicit scheme \\\midrule
        Fluids solver & Fractional projection & Pressure-velocity coupling \\
        ~ & $1^{st}$-order upwind/$2^{nd}$-order centred scheme  & Advection\\ 
        ~ &  Second-order central scheme & Diffusion \\
        ~ & CG/BiCG/GMRES & Pressure solver options
        \\ \midrule
        Solid solver & $4^{th}$-order Runge-Kutta & Rigid solid dynamics\\ 
        ~ & Newmark-$\beta$ & Flexible solid dynamics\\\midrule
        Subgrid model & Constant Smagorinsky model & For LES\\\midrule
        Wall model & Equilibrium wall shear stress model & For WMLES \\ \midrule
        Boundary conditions & Periodic & For periodic domains \\   
        ~ & Dirichlet & E.g., No-slip wall, velocity inlet \\ 
        ~ & Neumann & E.g., Free-slip wall, pressure outlet\\ 

    \bottomrule[1pt]
    \end{tabular} 
    \label{tab: boundary}  
    \end{table} 

\section{Validation of forward simulation capabilities} \label{s:validate}

We validate the accuracy, stability, and efficiency of Diff-FlowFSI across a series of forward simulation benchmarks. These tests are grouped into two categories:
\begin{itemize}
    \item \textbf{2D FSI benchmarks}: These cases focus on the solver's ability to handle strongly coupled fluid–structure interactions using the immersed boundary method. Scenarios include vortex shedding from a stationary obstacle, vortex-induced vibrations (VIV) of a rigid cylinder, and large deformation of flexible structures.
    \item \textbf{3D turbulent flow benchmarks}: These cases assess the solver's fidelity and performance in high-resolution eddy-resolving simulations (DNS, LES, WMLES) for flows involving wall-bounded turbulence, flow separation, and  complex vortex dynamics. 
\end{itemize}
Each case is designed to test different components of the solver — including the fluid solver, immersed boundary implementation, structural dynamics, and two-way coupling strategy — and to compare results against well-established data from the literature.

\subsection{Unsteady vortex shedding from a stationary cylinder} \label{s:sec4.1}

\begin{figure}[t!]
\centering
\includegraphics[width=\textwidth]{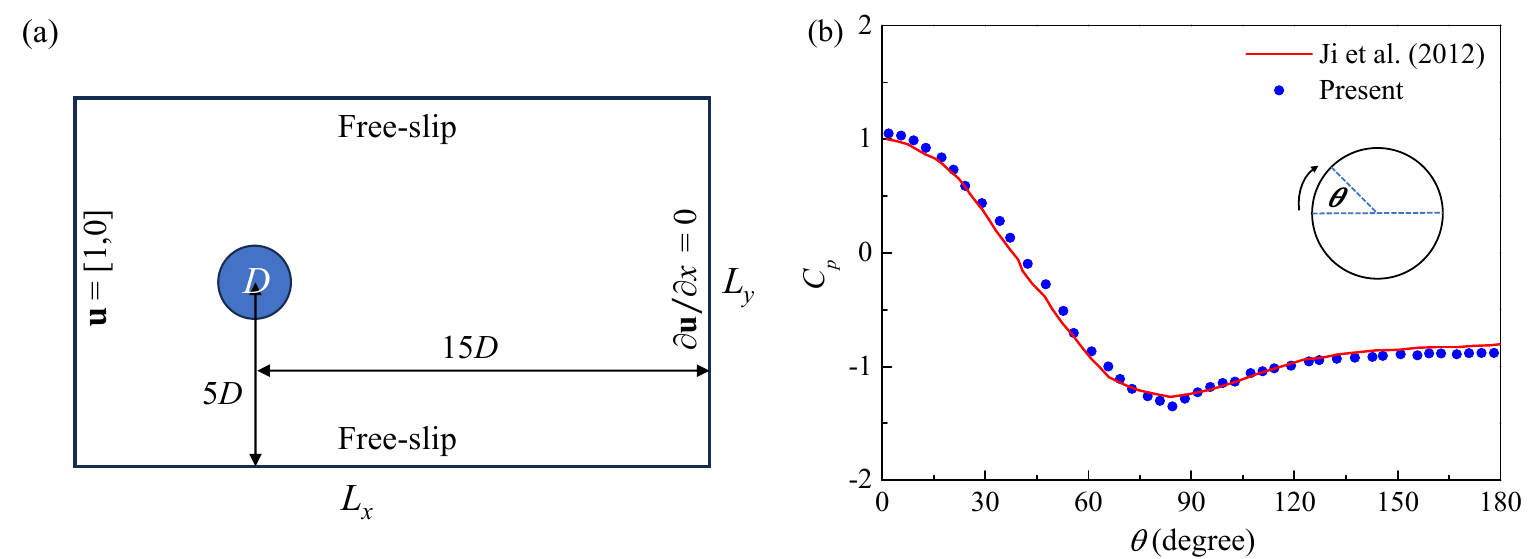}
\caption{(a) Computational domain; (b) pressure coefficients for static cylinder at $Re=100$. The reference data are taken from~\cite{ji2012novel}.}
\label{fig:cp_re100}
 \end{figure}

To validate the fluid solver and the implementation of immersed boundaries, we first simulate unsteady laminar vortex shedding from a stationary cylinder, an canonical benchmark in incompressible flow modeling. This case highlights the solver's ability to accurately resolve boundary layer separation, wake dynamics, and unsteady force coefficients at moderate Reynolds numbers. One key feature of Diff-FlowFSI is its flexibility in handling arbitrarily shaped immersed geometries. For demonstration, we consider a 2D cylinder with diameter $D$, placed in a rectangular domain of size $L_x \times L_y=20D \times 10D$, as shown in Figure~\ref{fig:cp_re100}(a). The computational grid consists of $N_x \times N_y = 512 \times 256$ uniformly spaced cells, and the simulation is advanced in time using a forward Euler scheme. The inflow boundary is prescribed using Dirichlet conditions with uniform velocity $\mathbf{u} = [u,v]=[1,0] $ m/s, while a Neumann outflow condition ${\partial \mathbf{u}}/{\partial x}=0$, and free-slip conditions are applied at the lateral boundaries. Simulations are run up to $tV/D=500$, ensuring that the flow has reached a statistically steady state.
Figure~\ref{fig:cp_re100}(b) presents the pressure coefficient $C_p$ along the cylinder surface at $Re=100$, showing excellent agreement with the reference solution of Ji et al~\cite{ji2012novel}.
\begin{figure}[H]
\centering
\includegraphics[width=\textwidth]{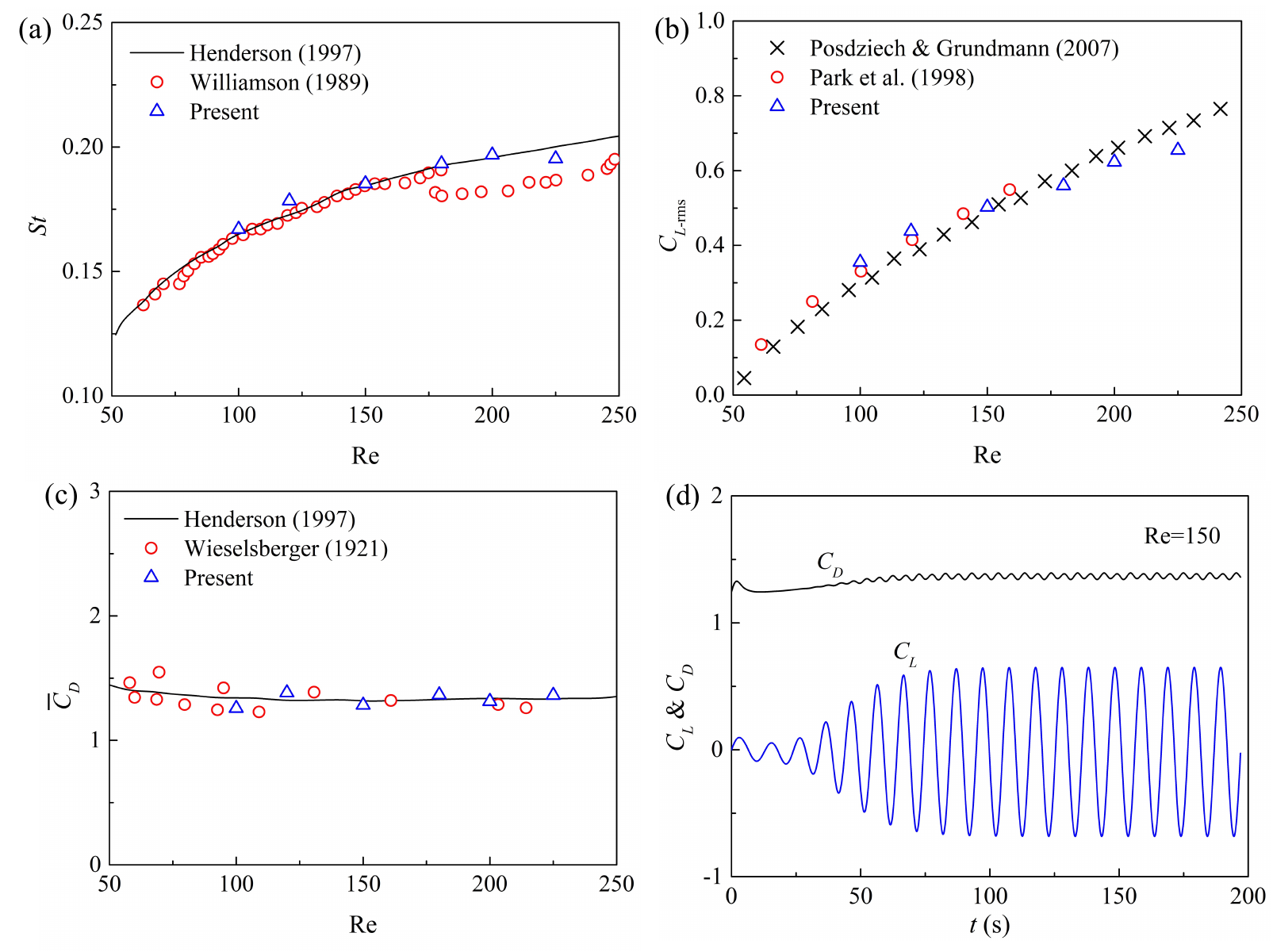}
\caption{Validation of unsteady vortex shedding from a stationary cylinder over the Reynolds number range $Re = 50\text{--}250$. (a) Strouhal number $St$; (b) RMS lift coefficient $C_{L,\text{rms}}$; (c) Mean drag coefficient $\overline{C}_D$; and (d) time histories of lift and drag coefficients at $Re = 150$. Reference data are taken from~\cite{henderson1997nonlinear, wieselsberger1921neuere, posdziech2007systematic, park1998numerical, williamson1989oblique}.}
\label{fig: static_cylinder_st_cl_cd}
 \end{figure}
To further validate the solver, we compute the Strouhal number $St$, root-mean-square lift coefficient $C_L$, and time-averaged drag coefficient $C_D$ across the Reynolds number range $Re \in [50, 250]$. As shown in Figure~\ref{fig: static_cylinder_st_cl_cd}(a)–(c), the results from Diff-FlowFSI closely match established reference data from experimental and numerical studies~\cite{henderson1997nonlinear, wieselsberger1921neuere, posdziech2007systematic, park1998numerical, williamson1989oblique}. Figure~\ref{fig: static_cylinder_st_cl_cd}(d) shows the time series of lift and drag forces at Re=150, capturing both the transitional and periodic stages of vortex shedding. 
\begin{figure}[t!]
\centering
\includegraphics[width=\textwidth]{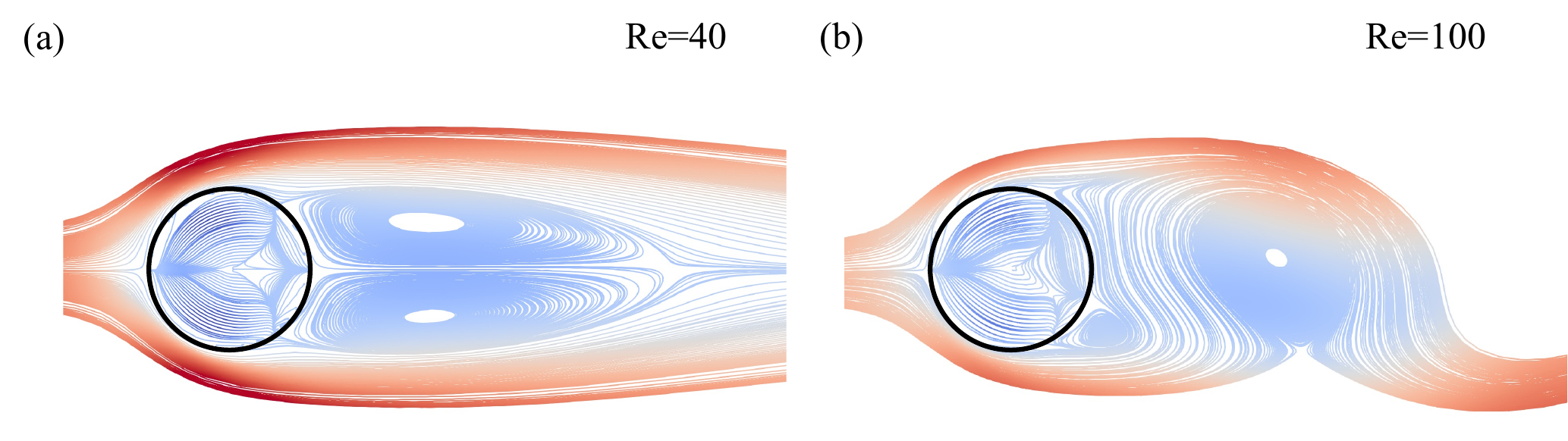}
\caption{Streamlines for vortex shedding from a static cylinder: (a) $Re=40$; (b) $Re=100$.}
\label{fig: vortex shedding}
 \end{figure}
Finally, we examine the time-averaged flow fields. At $Re = 40$, a steady recirculating bubble is observed in the wake, with a recirculation length of approximately $2D$, consistent with prior studies~\cite{park1998numerical} (Figure~\ref{fig: vortex shedding}(a)). At $Re = 100$, unsteady vortex shedding develops and the separation point and near-wake structure are accurately captured, as illustrated in Figure~\ref{fig: vortex shedding}(b), further validating the immersed boundary formulation and time integration scheme.

\subsection{Vortex-induced vibration (VIV) of a rigid cylinder} \label{s:sec4.2}

A rigid cylinder immersed in uniform flow and elastically supported exhibits vortex-induced vibration (VIV) due to periodic shedding of vortices. This canonical benchmark is used to evaluate the two-way FSI capabilities of Diff-FlowFSI. The computational domain and grid resolution match those described in Section~\ref{s:sec4.1}. Structural parameters used in this study are summarized in Table~\ref{tab:structural parameter}, consistent with prior studies~\cite{chen2018vortex, bao2012two}. The reduced velocity is defined as $U_r=\frac{V}{D f_n}$, where $V$ is the freestream velocity, $D$ is the cylinder diameter, and $f_n$ is the natural frequency of the structure. Simulations are performed over a range of reduced velocities $U_r \in [3,10]$, for both single-degree-of-freedom (1-DOF) and two-degree-of-freedom (2-DOF) configurations. 
The non-dimensional vibration amplitudes in the cross-flow and in-flow directions are defined as,
\begin{equation}
    A_y = \frac{\sqrt{2} y_{rms}}{D}, \ A_x=\frac{\sqrt{2} (x_{rms}-\overline{x})}{D}.
\end{equation}
where $y_{rms}$ and $x_{rms}$ are the root-mean-square (rms) displacements, and $\overline{x}$ is the time-averaged displacement in the streamwise direction.
\begin{table}[H]
    \centering
        \caption{Structural parameters for the elastically mounted cylinder}
    \begin{tabular}{c c c c}
        \toprule[1pt]
        Mass ration $m^*$ & Damping ratio $\xi_s$ & Natural frequency $f_n$ & $Re$ \\ \midrule
        2 & 0 & 0.1-0.3 & 100,~150 \\
        \bottomrule[1pt]
    \end{tabular}
    \label{tab:structural parameter}
\end{table}

\subsubsection{Single-degree-of-freedom (1-DOF) vibration response}

\begin{figure}[t!]
\centering
\includegraphics[width=\textwidth]{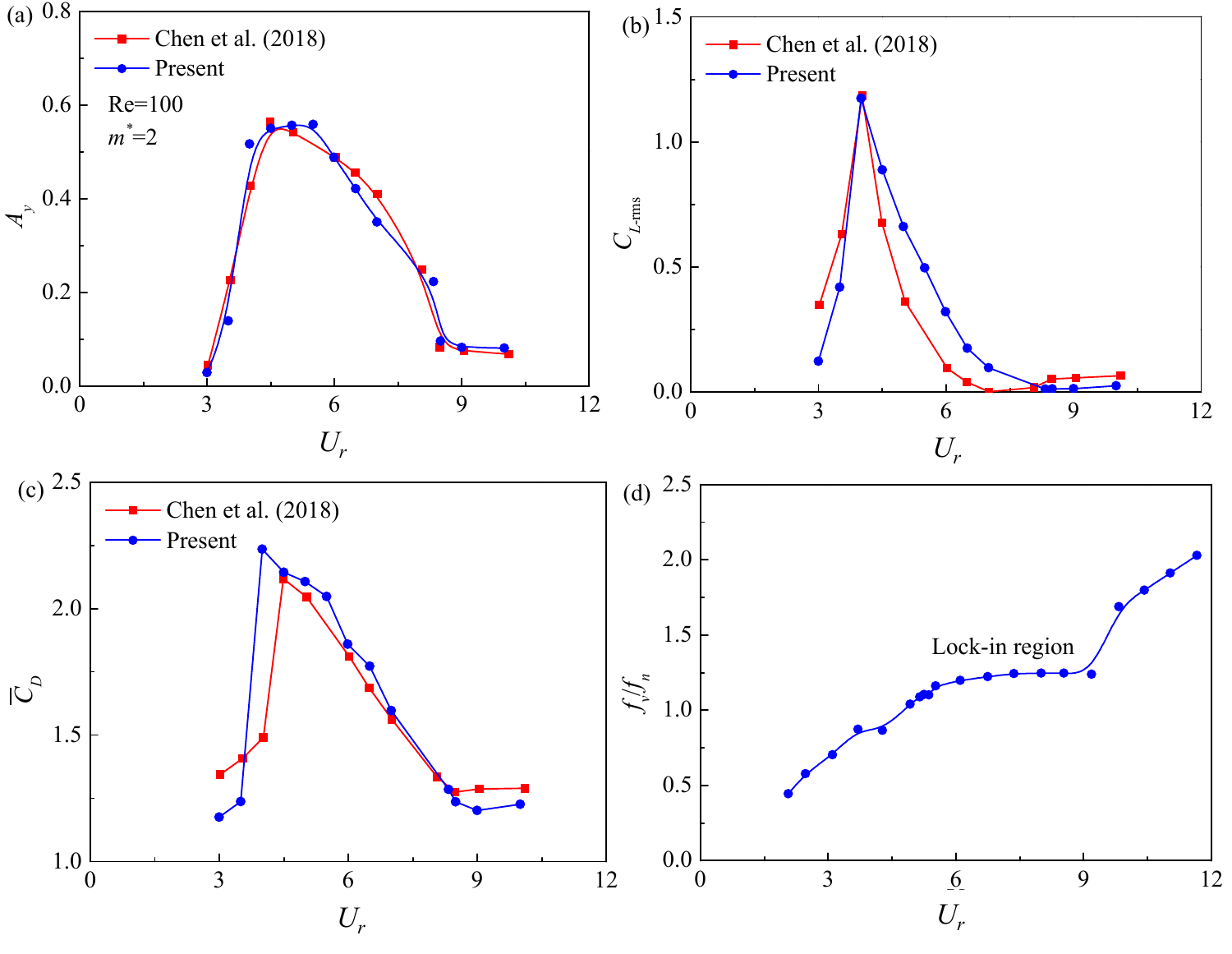}
\caption{Validation results for the 1-DOF VIV of a single cylinder: (a) cross-flow vibration amplitudes $A_{y}$; (b) rms lift force coefficient $C_{L}$;(c) mean drag force coefficient $\bar{C}_{D}$; and (d) the vortex shedding frequency to show the lock-in region.}
\label{fig: 1-DOF VIV}
 \end{figure}

Figure~\ref{fig: 1-DOF VIV} summarizes the 1-DOF VIV responses across a range of reduced velocities. Overall, the results show excellent agreement with the results reported in~\cite{chen2018vortex}. The cross-flow amplitudes $A_y$ (Figure~\ref{fig: 1-DOF VIV}(a)) exhibits the expected peak near resonance. The RMS lift coefficient $C_{L-\mathrm{rms}}$ and the mean drag coefficient $\overline{C}_D$ follow expected trends but show slight deviations at specific $U_r$, likely due to interpolation errors near the moving immersed boundary~\cite{gsell2021direct}. The lock-in region is identified within $U_r \in [6, 9]$, consistent with observations reported by~\cite{gabbai2005overview}. This frequency synchronization region is visualized in Figure~\ref{fig: 1-DOF VIV}(d).

\subsubsection{Two-degree-of-freedom (2-DOF) coupled oscillations} \label{s: 2dof}

\begin{figure}[t!]
\centering
\includegraphics[width=\textwidth]{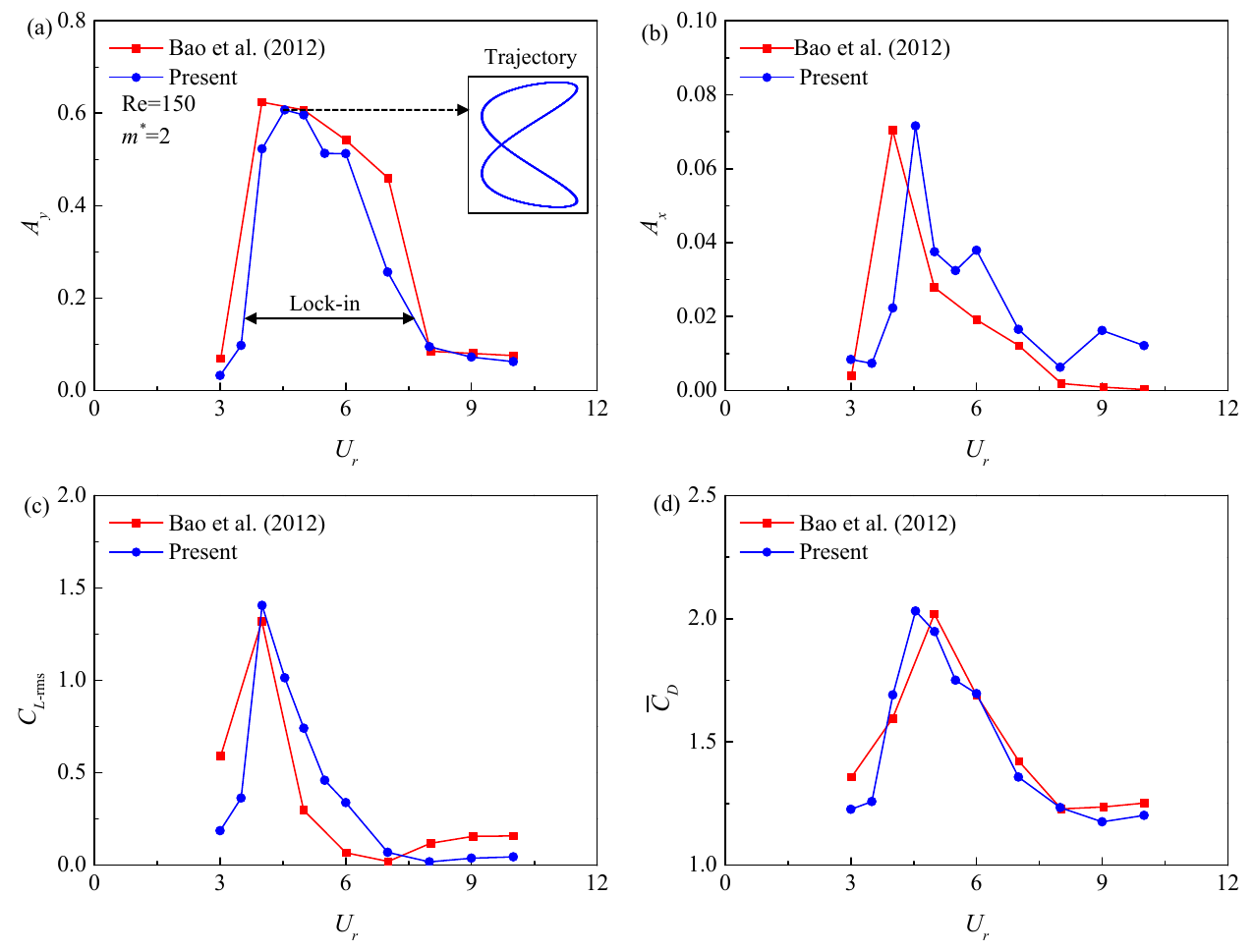}
\caption{Validation results for the 2-DOF VIV of a single cylinder: (a) cross-flow vibration amplitudes $A_{y}$; (b) in-flow vibration amplitudes $A_{x}$; (c) rms lift coefficient $C_{L}$; and (d) mean drag coefficient $C_{D}$.}
\label{fig: 2-DOF VIV}
 \end{figure}

Results for the 2-DOF elastically mounted cylinder are shown in Figure~\ref{fig: 2-DOF VIV}. Diff-FlowFSI accurately reproduces the amplitude responses and hydrodynamic forces across a range of reduced velocities $U_r \in [3, 10]$, demonstrating its robustness in capturing complex FSI phenomena. 
The cross-flow amplitude $A_y$ exhibits a pronounced peak within the lock-in region $U_r \in [5, 7]$, where the structural oscillation frequency synchronizes with the vortex shedding frequency. At $U_r = 5.5$, the vibration trajectory forms a characteristic figure-eight pattern, as shown in the inset of Figure~\ref{fig: 2-DOF VIV}(a), indicating that the shedding frequency in the cross-flow direction is approximately twice that in the in-flow direction. This harmonic synchronization reflects classic VIV dynamics and agrees with experimental observations in~\cite{srinil2013two}. The in-flow amplitude $A_x$ remain relatively small but follows a consistent trend with~\cite{bao2012two}. Some discrepancies in $C_{L,rms}$ and $\bar{C}_c$ are observed, particularly in the post-lock-in region, which may stem from differences in numerical schemes: the ALE formulation in~\cite{bao2012two} uses body-fitted grids and dynamic meshing, while Diff-FlowFSI adopts a sharp-interface immersed boundary method.  Despite these methodological differences, the overall agreement is strong, and both models capture the essential physics of the problem.

\begin{figure}[htp!]
\centering
\includegraphics[width=\textwidth]{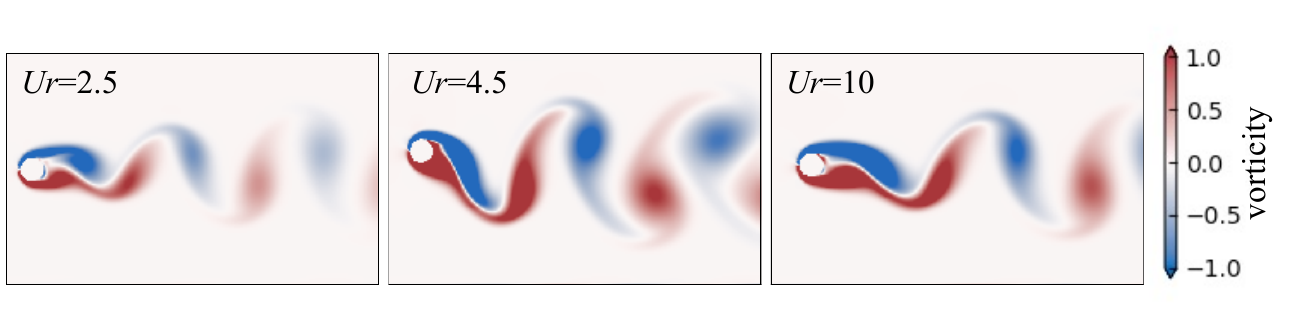}
\caption{Vorticity modes for the 2-DOF VIV of a single cylinder.}
\label{fig: 2-DOF vorticity}
 \end{figure}
The corresponding instantaneous vorticity fields in Figure~\ref{fig: 2-DOF vorticity} illustrate the transition between different vortex modes. At $U_r = 2.5$, the flow exhibits a symmetric ``2S'' shedding mode associated with the initial branch. At $U_r = 4.5$, a superposition of oscillatory motion and asymmetric vortex shedding defines the super branch, while at $U_r = 10$, a lower branch with weakened structural response and larger-scale vortex structures is recovered. These patterns are consistent with classical classifications of VIV shedding modes~\cite{prasanth2008vortex}, confirming the fidelity of the Diff-FlowFSI framework in resolving unsteady wake–structure interactions.

\subsection{Flow-induced deformation of flexible structures} \label{s:sec4.3}

This section demonstrates the capability of Diff-FlowFSI to simulate large-deformation dynamics of flexible bodies undergoing strong fluid–structure interactions. Two canonical benchmarks are selected: (i) a vertical flexible plate anchored at its base, and (ii) a flexible plate attached to the leeside of a fixed circular cylinder. These test cases capture different coupling regimes, including static reconfiguration and self-excited oscillations driven by unsteady wake–structure interaction. The computational setup for both cases (domain size, grid resolution, boundary conditions) follows the configuration described in Section~\ref{s:sec4.1}. Structural parameters including elasticity, damping, and density ratios are summarized in Table~\ref{tab:flexible parameter}. Due to the low stiffness of the structures, strong added-mass effects are present. To ensure numerical stability and accuracy, we employ the strong coupling strategy described in Algorithm~\ref{scheme: strong-coupling}, with a convergence tolerance of $\xi_t = 10^{-5}$. On average, each time step requires 2–3 sub-iterations for convergence.
\begin{table}[H]
    \centering
    \small
    \caption{Structural parameters for flexible plates}
    \begin{tabular}{c c c c c c}
    \toprule
      Cases & Young's modulus $E$/Pa  & Poisson’s ratio & damping ratio $\xi_s$ & density ratio $\rho*$ & $Re$\\ \midrule
       Vertical &  $5\times10^4$ & 0.3 &0.001 &10 & 600 \\
       Attached &  $4\times10^3$ & 0.3 &0.001 &5 & 500 \\ 
    \bottomrule
    \end{tabular}
    \label{tab:flexible parameter}
\end{table}

\subsubsection{Static reconfiguration of a vertical flexible plate}

A vertically oriented flexible plate clamped at its base is subjected to a steady inflow. This canonical configuration serves as a simplified model for understanding drag reduction and reconfiguration in aquatic vegetation and other bioinspired systems. The key dimensionless parameter governing deformation is the rigidity number $\lambda$, defined as:
\begin{equation}
\lambda = \frac{EI}{\rho V ^2 L^3} .
\label{eq:rigidity}
\end{equation}
where $EI$ is bending stiffness, $V$ is the freestream velocity, and $L$ is the plate length. For the current configuration, $\lambda = 0.15$, which falls within the static reconfiguration regime~\cite{zhang2020fluid}.

\begin{figure}[htp!]
\centering
\includegraphics[width=\textwidth]{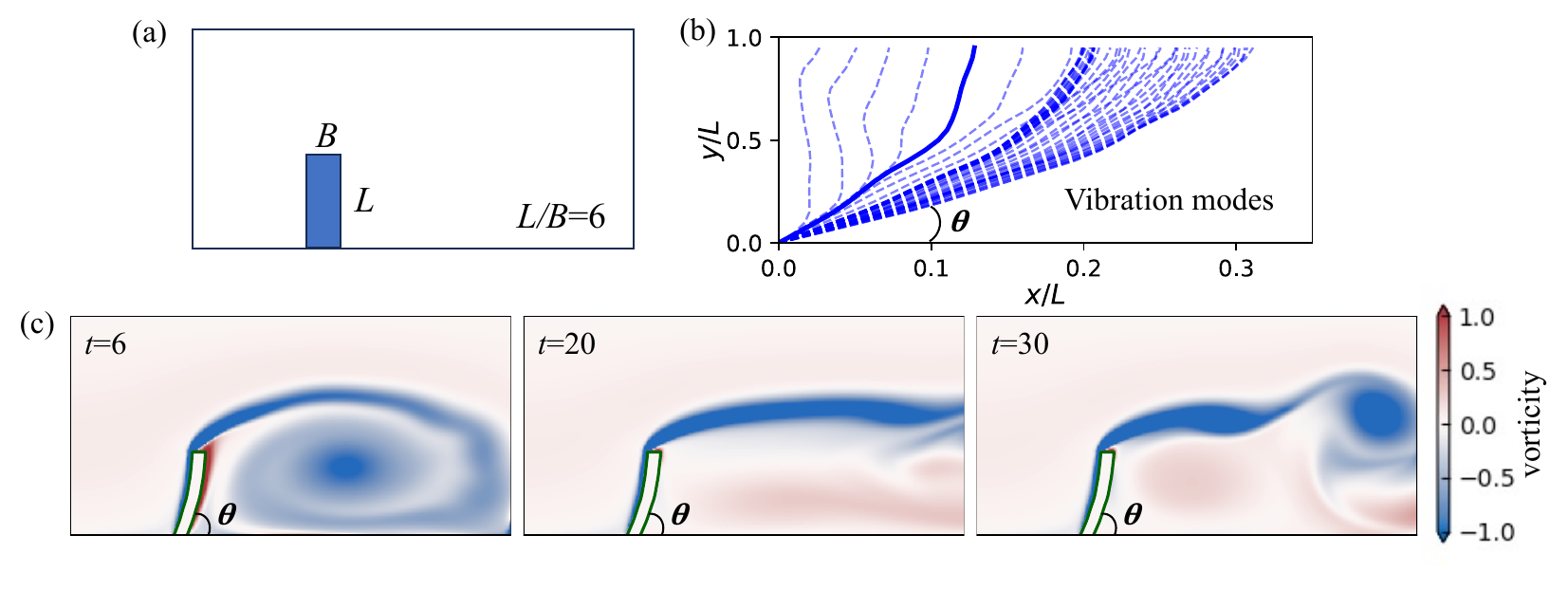}
\caption{Validation of static reconfiguration of a vertical flexible plate: (a) computational domain and boundary setup; (b) deformed equilibrium shape and chord line angle; (c) instantaneous vorticity field revealing wake recirculation.}
\label{fig: vertical_flexible}
 \end{figure}

As shown in Figure~\ref{fig: vertical_flexible}, the plate deforms steadily under fluid loading and reaches an equilibrium configuration. The inclination angle of the chord line, defined between the root and the tip, is measured to be $\theta \approx 72^{\circ}$, which is in excellent agreement with the experimental measurements reported in~\cite{zhang2020fluid}. The instantaneous vorticity field in Figure~\ref{fig: vertical_flexible}(c) shows a well-developed recirculation region in the wake, confirming that the solver accurately captures the steady wake–structure interaction.

\subsubsection{Self-excited oscillation of a cylinder-mounted flexible plate}

\begin{figure}[htp!]
\centering
\includegraphics[width=\textwidth]{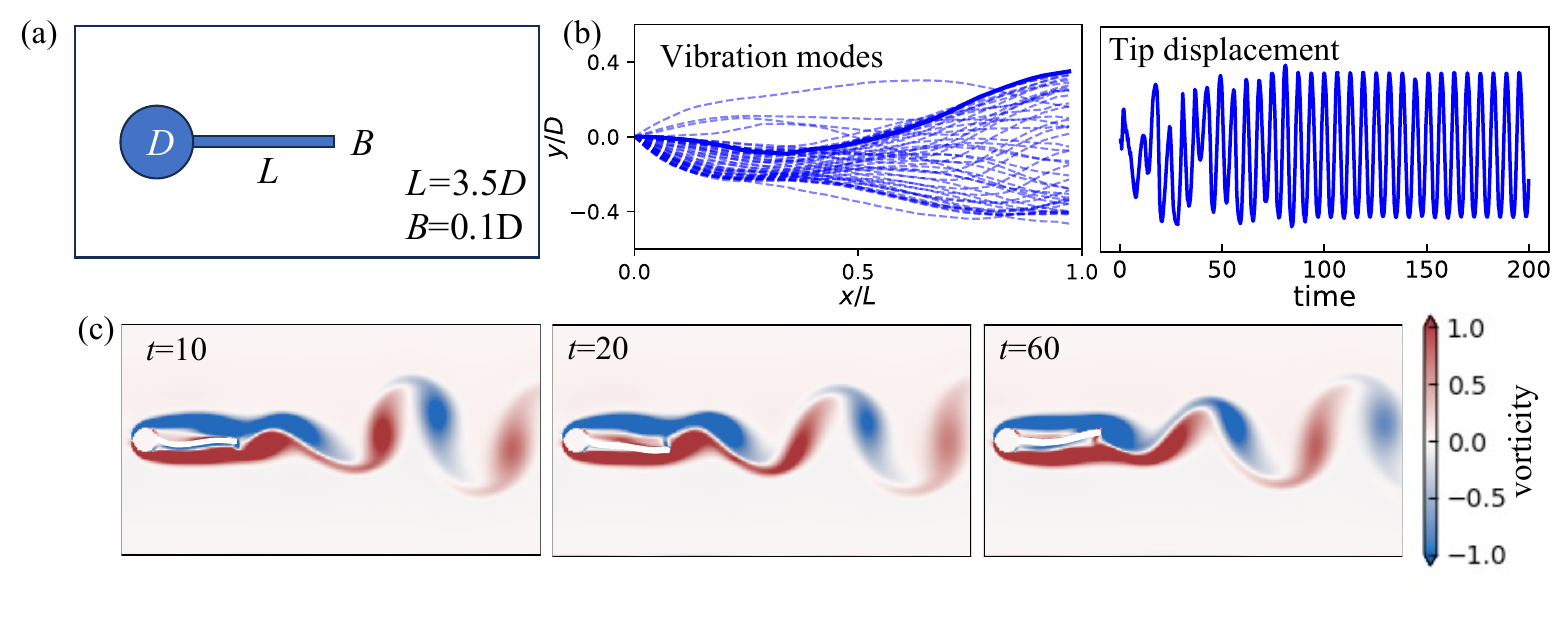}
\caption{Validation of oscillatory response of a flexible plate mounted on a cylinder: (a) computational domain and setup; (b) tip displacement trajectory and vibration shape; (c) instantaneous vorticity field revealing wake–plate interaction.}
\label{fig: cylinder_flexible}
 \end{figure}

In this benchmark, a flexible plate is attached to the downstream face of a fixed circular cylinder. This configuration introduces a strong interaction between the plate's elastic response and the unsteady vortex shedding from the upstream cylinder (see Figure~\ref{fig: cylinder_flexible}(a)).

As shown in Figure~\ref{fig: cylinder_flexible}(b), the plate undergoes self-sustained periodic oscillations. The amplitude of the tip displacement reaches $y_{\mathrm{tip}}/D \approx 0.3$, and the temporal response exhibits regular, nearly sinusoidal behavior. These results agree well with the observations in~\cite{pfister2020fluid}, which reported similar periodic flapping and deformation patterns for flexible plates of comparable stiffness. Figure~\ref{fig: cylinder_flexible}(c) shows the instantaneous vorticity field, where a symmetric ``2S'' vortex shedding pattern dominates the near-wake dynamics, again matching prior findings~\cite{pfister2020fluid}. These results demonstrate that Diff-FlowFSI can robustly simulate unsteady FSI phenomena involving low-rigidity structures coupled with complex vortex dynamics.

\subsection{Three-dimensional wall-bounded turbulence} \label{s:turbulence}

This section evaluates the capability of Diff-FlowFSI to resolve wall-bounded turbulence in three-dimensional (3D) settings. We consider several canonical benchmarks to assess accuracy, robustness, and applicability to simulating inhomogeneous, anisotropic turbulent flows with complex wall geometries. These tests include: (i) turbulent channel flow at friction Reynolds number $Re_\tau = 180$; (ii) channel flow over rough walls represented by immersed cubes; and (iii) separated flow over periodic hills. These cases also serve to validate the accuracy of the implemented Smagorinsky subgrid-scale (SGS) model and the equilibrium wall model, and to demonstrate the effectiveness of the IBM in representing complex geometries.

\subsubsection{Turbulent channel flow with smooth walls}

We begin with a well-established benchmark: fully developed turbulent channel flow at friction Reynolds number $Re_\tau = 180$, defined as $Re_\tau = u_\tau\delta/\nu$, where $u_\tau$ is the friction velocity, $\nu$ is the kinematic viscosity and $\delta$ is the channel half-height. The computational domain, grid resolution, and simulation parameters are listed in Table~\ref{tab:setup for 3D channel}. Streamwise, wall-normal, and spanwise directions are denoted by $x, y$, and $z$, respectively. The superscript $\left\langle + \right\rangle$ denotes normalization in wall units. Specifically, spatial variables are normalized by $\nu / u_{\tau}$, temporal variables are scaled by $\nu / u_{\tau}^2$, and velocity components are normalized by $u_{\tau}$.  Periodic boundary conditions are imposed in the $x$ and $z$ directions, and no-slip conditions are enforced on the top and bottom walls. The flow is driven by a constant pressure gradient in the streamwise direction. The forward Euler method is used for time integration with a fixed time step $\Delta t^+$, and the simulation is run for 20 flow-through times. Turbulent statistics are collected over the final 3 flow-throughs.
\begin{table}[H]
    \centering
    \small
    \caption{Numerical setup for the 3D turbulent channel flow at $Re_\tau = 180$}
    \begin{tabular}{c c c c c c c}
    \toprule
       Domain ($L_x \times L_y \times L_z$)  & Cell ($N_x \times N_y \times N_z$) & $\Delta x^+$ & $\Delta z^+$ & $\Delta y^+$ & $\Delta t^+$ & $T_{\mathrm{flow}}$\\ \midrule
        $2\pi \times 2 \times \pi$ & $160 \times 400 \times 100$ & 7.07 & 5.65 & 0.9 & $4 \times 10^{-3}$ & 22\\ 
    \bottomrule
    \end{tabular}
    \label{tab:setup for 3D channel}
\end{table}

\begin{figure}[htp!]
\centering
\includegraphics[width=\textwidth]{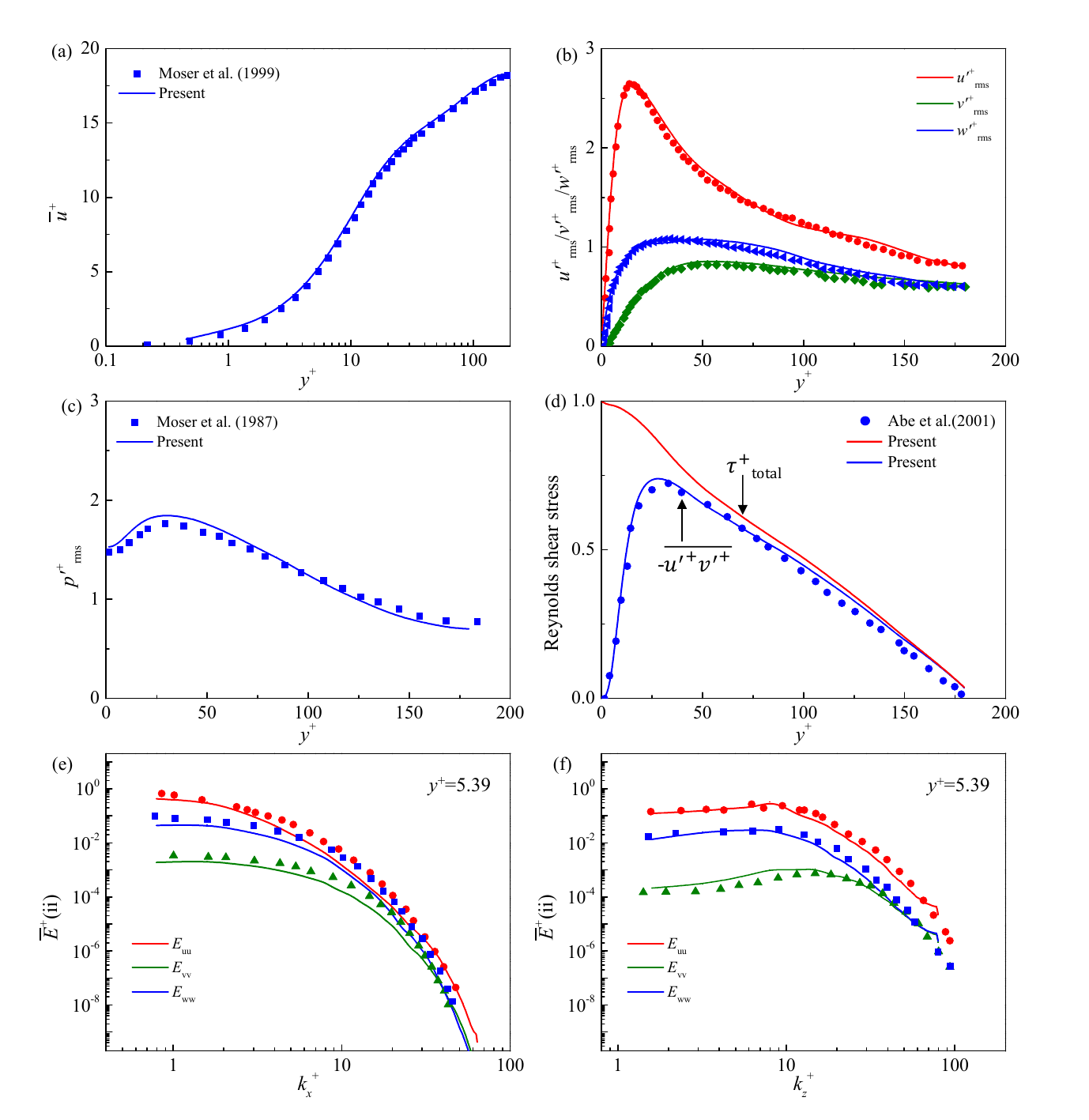}
\caption{Validation results for 3D turbulent channel flow at $Re_\tau = 180$: (a) normalized mean streamwise velocity profile $u^+$ compared against DNS data from~\cite{moser1999direct}; (b) turbulence intensities $u'_{\mathrm{rms}}$, $v'_{\mathrm{rms}}$, and $w'_{\mathrm{rms}}$, with reference data from~\cite{moser1999direct}; (c) root-mean-square wall pressure fluctuations $p'_{\mathrm{rms}}$, with DNS data from~\cite{kim1987turbulence}; (d) Reynolds shear stress and total shear stress profiles, compared with data from~\cite{abe2001direct}; (e) premultiplied energy spectra in the streamwise direction, and (f) in the spanwise direction, both evaluated at $y^+ \approx 5.4$ and benchmarked against~\cite{rai1991direct}.}
\label{fig:turbulence-statistic}
\end{figure}

Figure~\ref{fig:turbulence-statistic} presents a comprehensive comparison of turbulence statistics computed by Diff-FlowFSI against reference DNS datasets from Moser et al.\cite{moser1999direct}, Kim et al.\cite{kim1987turbulence}, and Abe et al.~\cite{abe2001direct}. Overall, the results demonstrate excellent agreement with the reference data. 
Panel (a) shows the mean streamwise velocity profile in wall units ($u^+ = \bar{u}/u_\tau$ vs. $y^+ = yu_\tau/\nu$). The logarithmic and viscous sublayer regions are well resolved, and the predicted profile matches the DNS reference data~\cite{moser1999direct} across the entire wall-normal extent. The observed friction coefficient $C_f=8.0 \times 10^{-3}$ lies within the expected range based on empirical correlations~\cite{dean1978reynolds, steen2017saph}. Panel (b) plots the RMS velocity fluctuations in all three directions normalized by $u_\tau$. Diff-FlowFSI accurately captures both the peak near-wall fluctuations and their decay toward the channel centerline. The peak in streamwise fluctuations ($u'^+_{rms} \approx 2.7$) and location of the peak ($y^+ \approx 15$) are consistent with DNS data. Panel (c) presents the RMS wall pressure fluctuations, which are notoriously sensitive to numerical schemes and resolution. The results from Diff-FlowFSI align closely with the DNS data from Kim et al.~\cite{kim1987turbulence}, including the peak location and magnitude. Panel (d) shows the Reynolds shear stress profile $-\overline{u'^+v'^+}$ alongside the total shear stress $\tau_\text{total}^+$. The intersection point of the turbulent and viscous contributions correctly occurs at $y^+ \approx 12$, and the profile conforms well to the expected linear decay, validating the physical consistency of the flow field. Panels (e) and (f) plot the one-dimensional energy spectra in the streamwise and spanwise directions at $y^+ = 5.39$. The spectra show the expected decay and match the DNS results from Rai and Moin~\cite{rai1991direct}, confirming the resolution of both large and small turbulent structures. Notably, the flattening and tail behavior at high wavenumbers are well reproduced, indicating sufficient spatial resolution and low numerical dissipation.

\begin{figure}[htp!]
\centering
\includegraphics[width=\textwidth]{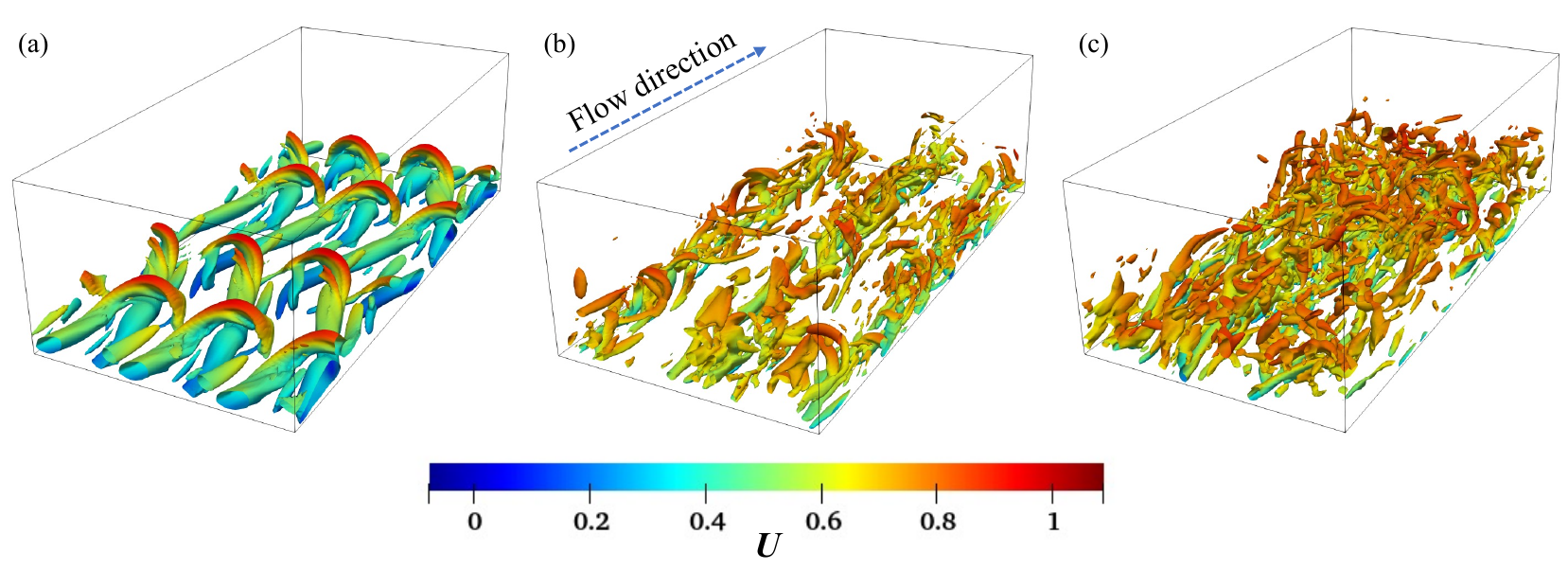}
\caption{The vortical structures for 3D turbulence based on Q criteria at different flow through time, where only half channel is presented: (a) $T_{\mathrm{flow}}=1$ (b) $T_{\mathrm{flow}}=14$ and (c) $T_{\mathrm{flow}}=20$.}
\label{fig:turbulence-vortical}
 \end{figure}

The evolution of vortical structures in the lower half of the channel is visualized in Figure~\ref{fig:turbulence-vortical} using iso-surfaces of the Q criterion, colored by normalized streamwise velocity magnitude. 
At the early stage ($T_{\mathrm{flow}} = 1$), the flow exhibits a pattern of quasi-regular streamwise-aligned vortices near the wall, indicative of laminar-to-transitional dynamics seeded by initial perturbations. By $T_{\mathrm{flow}} = 14$, these coherent structures begin to break down, forming elongated streaks and 3D instabilities. At $T_{\mathrm{flow}} = 20$, the flow field has transitioned to fully-developed turbulence, characterized by a dense population of fine-scale, anisotropic vortices distributed throughout the near-wall region and extending into the channel center. This progression illustrates the solver's ability to capture key physical features of wall-bounded turbulent transition and sustain a statistically stationary turbulent regime over long temporal integrations.

\subsubsection{Turbulent boundary layer with wall roughness}

To evaluate the capability of Diff-FlowFSI in simulating turbulent flows over rough surfaces, we perform a DNS of a rough-wall turbulent boundary layers with uniformly distributed cubic roughness elements. Starting from a statistically steady turbulent channel flow at $Re_\tau = 180$, the upper wall is replaced by a free-slip Neumann boundary, and a layer of immersed cubic obstacles is introduced along the bottom wall. The height of each cube is denoted by $h$, which serves as the characteristic length scale in this configuration. The Reynolds number based on the top mean velocity $U$ and cube height is defined as $Re = Uh/\nu = 3200$. The computational domain is set to $L_x \times L_y \times L_z = 9h \times 6h \times 6h$, discretized with a high-resolution grid of $200 \times 400 \times 300$ cells to adequately resolve flow structures within and above the roughness layer. 

\begin{figure}[t!]
\centering
\includegraphics[width=1.0\textwidth]{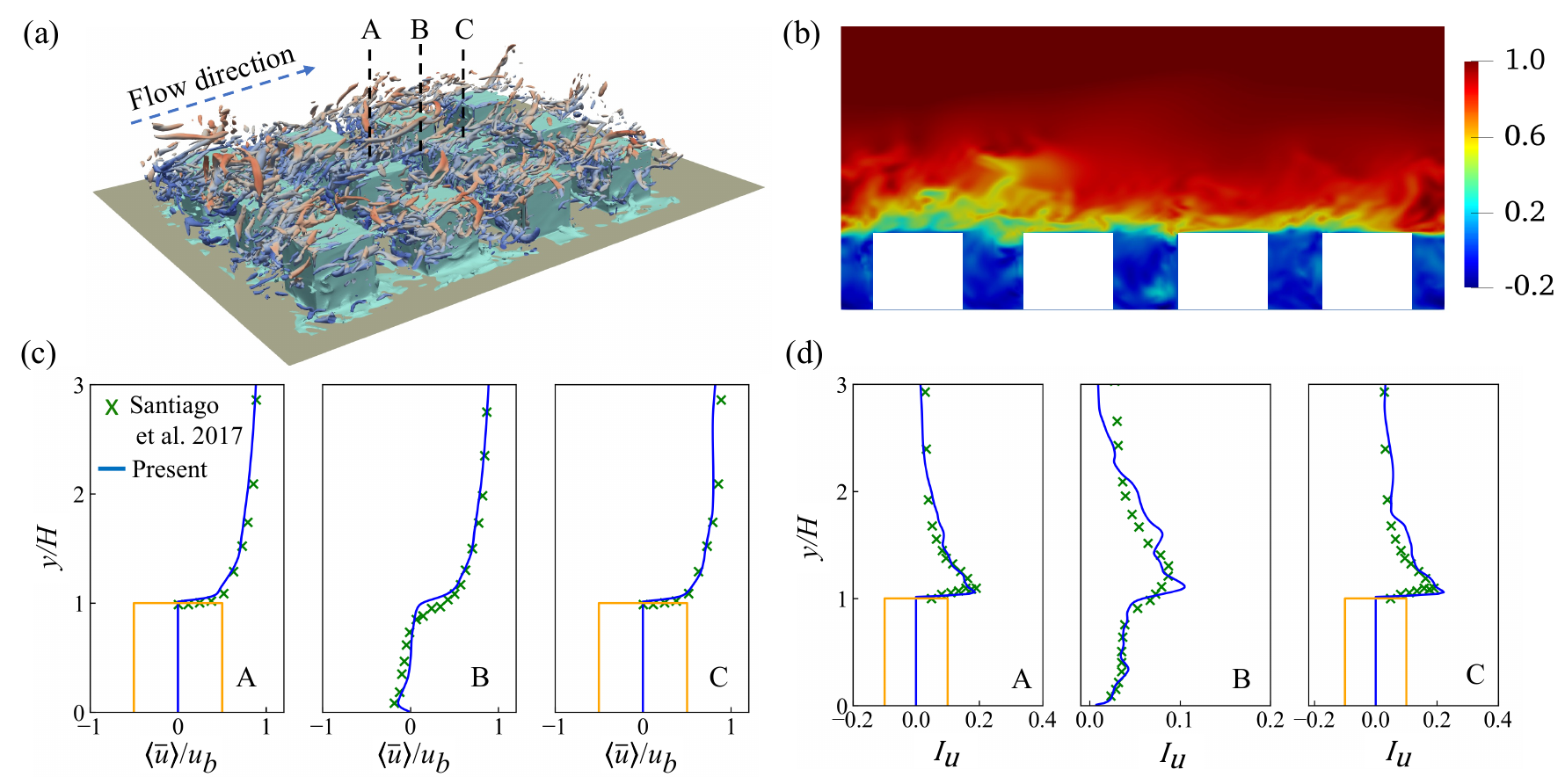}
\caption{Validation results for turbulent flow over cubic rough walls: (a) instantaneous 3D vortical structures visualized by Q-criterion iso-surfaces; (b) streamwise velocity contours in the vertical ($x-y$) plane; (c) mean streamwise velocity profile $\bar{u}/u_b$, where $u_{b}$ is the bulk velocity; and (d) streamwise turbulence intensity $I_u = u'_\text{rms}/u_b$. Reference data are taken from Santiago et al.~\cite{santiago2007cfd}.}
\label{fig:3Dcube-contour}
 \end{figure}

Figure~\ref{fig:3Dcube-contour}(a) shows the instantaneous vortical structures visualized using Q-criterion iso-surfaces, highlighting the complex vortex shedding and turbulent mixing generated by the cubic roughness. The streamwise velocity contours in the vertical plane, as shown in Panel (b), further illustrate the separation, recirculation, and wake dynamics behind each obstacle.
Quantitative comparisons are shown in Figures~\ref{fig:3Dcube-contour}(c) and (d), where mean velocity profiles and turbulence intensity ($I_u$) are evaluated at three representative streamwise locations (A: over a cube, B: within a canyon, and C: over another cube). Results are benchmarked against the experimental and numerical data of Santiago et al.~\cite{santiago2007cfd}. As shown in Figure~\ref{fig:3Dcube-contour}(c), the predicted mean velocity profiles closely match the reference data at all three locations, validating the solver's ability to resolve flow acceleration above the roughness and low-speed recirculation within the canyons.
Figure~\ref{fig:3Dcube-contour}(d) presents the corresponding turbulence intensity profiles. While the simulation reproduces the general shape and peak location of the intensity distribution, slight overprediction is observed in the upper layers, particularly at location C. This discrepancy is likely attributed to the sharp-interface IBM’s interpolation error near the obstacle boundary, which can influence shear-layer development and turbulence generation. Potential improvements include refined marker placement, reduced time step size, or incorporation of penalty-based stabilization~\cite{cai2017moving, zhou2021analysis, kim2010simulating}.
 
\subsubsection{Separated turbulent flow over periodic hills}\label{s:periodic_hill}

To further assess the capability of Diff-FlowFSI in capturing separated turbulent flows over complex geometries, we simulate the canonical periodic hill problem~\cite{breuer2009flow}. This benchmark presents significant challenges due to flow separation, reattachment, adverse pressure gradients, and the presence of large-scale coherent structures, making it a stringent test for numerical solvers. The computational setup follows the geometry defined in~\cite{breuer2009flow}, with periodic boundary conditions imposed in the streamwise and spanwise directions. No-slip conditions are applied at the top and bottom walls. The Reynolds number based on the hill height $h$ and bulk velocity $u_b$ is Re = 2800. The domain is discretized using a structured grid with refined resolution near the wall and in regions of flow separation and recirculation.

 \begin{figure}[t!]
\centering
\includegraphics[width=\textwidth]{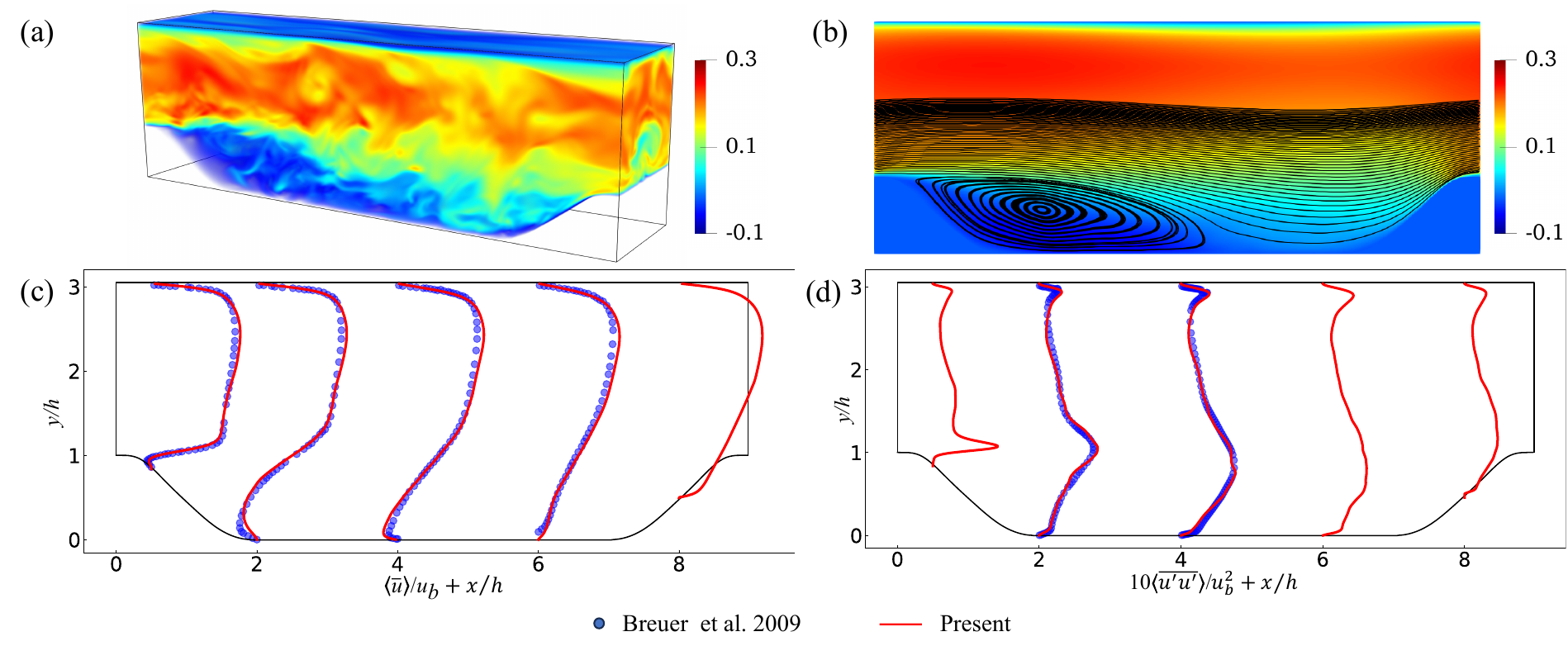}
\caption{Validation results for separated turbulent flow over periodic hills: (a) instantaneous 3D velocity contours; (b) time-averaged streamwise velocity with overlaid streamlines, highlighting the recirculation region; (c) mean streamwise velocity profiles $\overline{u}/u_b$, where $u_b$ is the bulk velocity; (d) streamwise Reynolds normal stress $\overline{u'u'}$. Reference data are taken from~\cite{breuer2009flow}. The Reynolds number based on hill height and bulk velocity is $Re = 2800$.}
\label{fig:periodichill}
 \end{figure}
Figure~\ref{fig:periodichill}(a) displays the instantaneous three-dimensional velocity contours, revealing rich turbulent structures and large-scale flow separation downstream of the hills. Panel (b) shows the time-averaged streamwise velocity field overlaid with streamlines, highlighting the separation bubble and reattachment region. The size, shape, and location of the recirculation zone agree closely with those reported in~\cite{breuer2009flow}, demonstrating the fidelity of the immersed boundary treatment and solver stability. Statistical quantities are shown in Figures~\ref{fig:periodichill}(c)–(d). The mean streamwise velocity profiles $\overline{u}/u_b$ at five streamwise locations are plotted in panel (c), showing excellent agreement with the benchmark DNS data. The model successfully captures the velocity deficit within the recirculation region and the recovery downstream. In panel (d), the streamwise Reynolds stress $\overline{u{'}u{'}}$ is evaluated at the same locations. The peak turbulence intensities and their vertical locations match closely with reference, confirming that both mean and fluctuating components of flow are well resolved.

\subsubsection{Turbulent wake of a finite-length cylinder}

\begin{figure}[t!]
\centering
\includegraphics[width=1.0\textwidth]{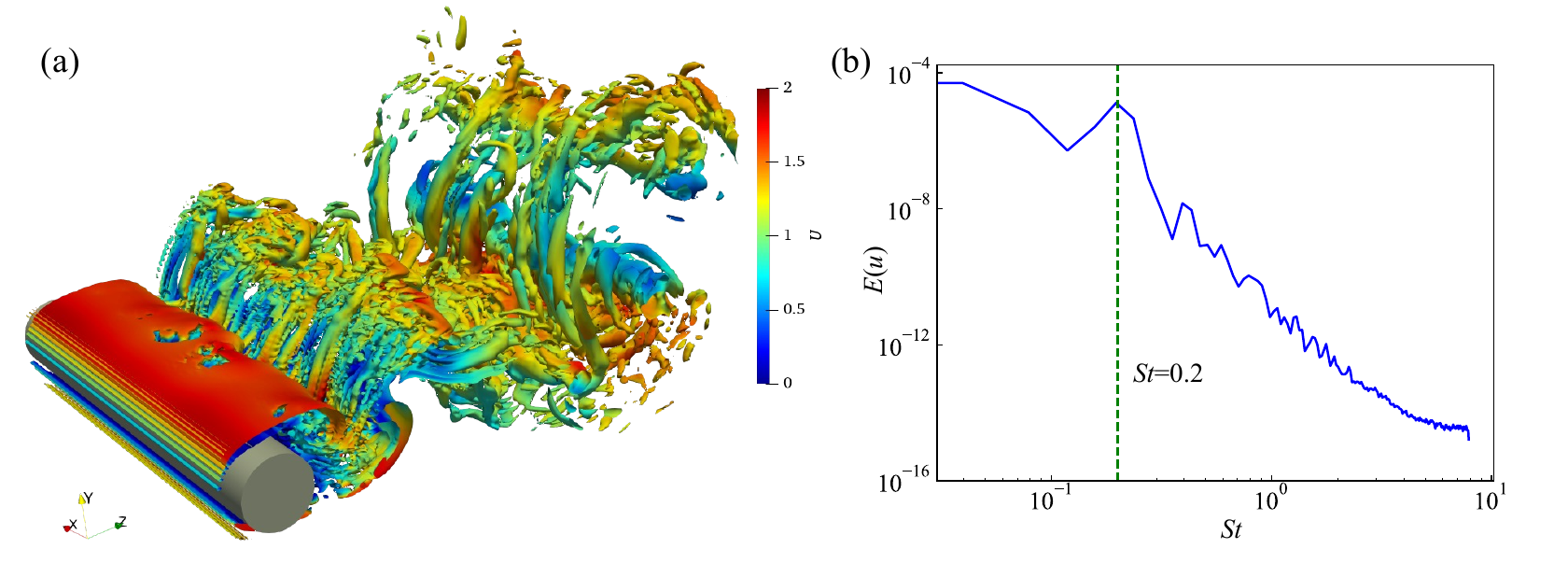}
\caption{Validation results for turbulent vortex shedding behind a 3D finite-length cylinder at $Re = 1000$: (a) instantaneous vortical structures extracted using the Q criterion; (b) power spectral density of the lift force showing the dominant shedding frequency at $St \approx 0.2$.}
\label{fig:3DVIV-contour}
\end{figure}
This test case evaluates Diff-FlowFSI's ability to simulate turbulent wake dynamics behind bluff bodies, emphasizing vortex shedding and non-equilibrium flow structures in the wake region. It serves as a 3D extension of the 2D benchmark discussed in Sec.~\ref{s:sec4.1}. The computational setup retains the same inflow velocity, domain length, and boundary conditions in the streamwise and cross-stream directions, but introduces a finite-length cylinder aligned along the spanwise ($z$) direction. The spanwise domain extends $L_z = 7D$, where $D$ is the cylinder diameter, and is discretized into $N_z = 100$ grid points. Periodic boundary conditions are imposed in the spanwise direction to mimic an infinite cylinder while maintaining computational efficiency. The simulation is performed at Reynolds number $Re = 1000$, based on the inflow velocity and cylinder diameter, placing the flow in the subcritical vortex shedding regime. A fully developed wake is established over several flow-through times using a forward Euler time integration scheme. 

Figure~\ref{fig:3DVIV-contour}(a) shows an instantaneous snapshot of the vortical structures extracted using the Q-criterion. Characteristic alternating vortex rollers are clearly observed, forming a 3D von Kármán vortex street in the near wake. In panel (b), the temporal power spectral density (PSD) of the lift force signal is plotted against the Strouhal number, revealing a dominant peak at $St \approx 0.2$. This value is in close agreement with empirical and numerical benchmarks for circular cylinders in this Reynolds number regime~\cite{douglas2022strouhal}, confirming the accuracy of the unsteady vortex dynamics resolved by Diff-FlowFSI.

 \subsubsection{Wall-modeled LES of high-Re turbulent channel flows}
 
In addition to DNS capabilities, Diff-FlowFSI supports LES through the implementation of a constant-coefficient Smagorinsky subgrid-scale (SGS) model~\cite{scotti1993generalized}. While straightforward to implement, the constant SGS model is known to underpredict near-wall shear stress with relatively coarse, uniform grids. To address this limitation, Diff-FlowFSI incorporates an equilibrium wall model based on the Spalding law~\cite{spalding1961single}, which provides an algebraic closure for the instantaneous wall shear stress as a function of the resolved near-wall velocity. Specifically, the wall model is formulated as~\cite{bae2021effect, liefvendahl2017formulation} 
\begin{equation}
 u^+ = y^+ - e^{-\kappa B} \left[ e^{\kappa u^+} -1 - \kappa u^+ -\frac{1}{2} (\kappa u^+)^2 -\frac{1}{6} (\kappa u^+)^3\right]   ,
 \label{eq: wall model}
\end{equation}
where $u^+=u/u_{\tau}$ is instantaneous LES velocity scaled by the friction velocity, and the constants $\kappa = 0.4$ and $B = 5$ follow standard calibration. The equation is solved numerically using a Newton–Raphson root-finding method to obtain the local friction velocity $u_\tau$, which then yields the wall shear stress $\tau_w = \rho u_\tau^2$. To enforce the modeled wall shear stress in the simulation, the SGS viscosity at the first off-wall grid points is modified as:
\begin{equation}
\nu_{ti}=\frac{\tau_{wi}}{\frac{du_i}{dy}+\varepsilon}-\nu,
\label{eq:modify-sgs}
\end{equation}
where $i=1,3$ corresponds to the streamwise direction $x$ and spanwise direction $z$, respectively; $\varepsilon = 1\times10^{-6}$ is used prevent division by zero.  The directional components of the wall shear stress are scaled as:
\begin{equation}
\tau_{w1}=\tau_w \frac{u}{\sqrt{u^2+w^2}} , ~~\\
 \tau_{w3}=\tau_w \frac{w}{\sqrt{u^2+w^2}},
 \label{eq:modify-sgs-two-components}
\end{equation}
ensuring consistent tangential stress distribution at the wall.

We apply this wall-modeled LES (WMLES) framework to simulate turbulent channel flows over a wide range of friction Reynolds numbers, $Re_\tau = 180$ to $10^4$. The computational setup for all cases is summarized in Table~\ref{tab:wmles}. All simulations are conducted using a coarse grid with $64^3$ points, enabling fast evaluation of the wall model under limited resolution.
\begin{table}[htp!]
    \centering
    \begin{tabular}{c c c}
    \toprule
        Domain ($L_x \times L_y \times L_z$) & Grid ($N_x \times N_y \times N_z$) & $T_{\mathrm{flow}}$ \\ \midrule
       $2 \pi \times 2 \times \pi$  & $64 \times 64 \times 64$ & 10 \\
    \bottomrule
    \end{tabular}
    \caption{The computational settings for the cases in WMLES}
    \label{tab:wmles}
\end{table}

\begin{figure}[htp!]
\centering
\includegraphics[width=0.6\textwidth]{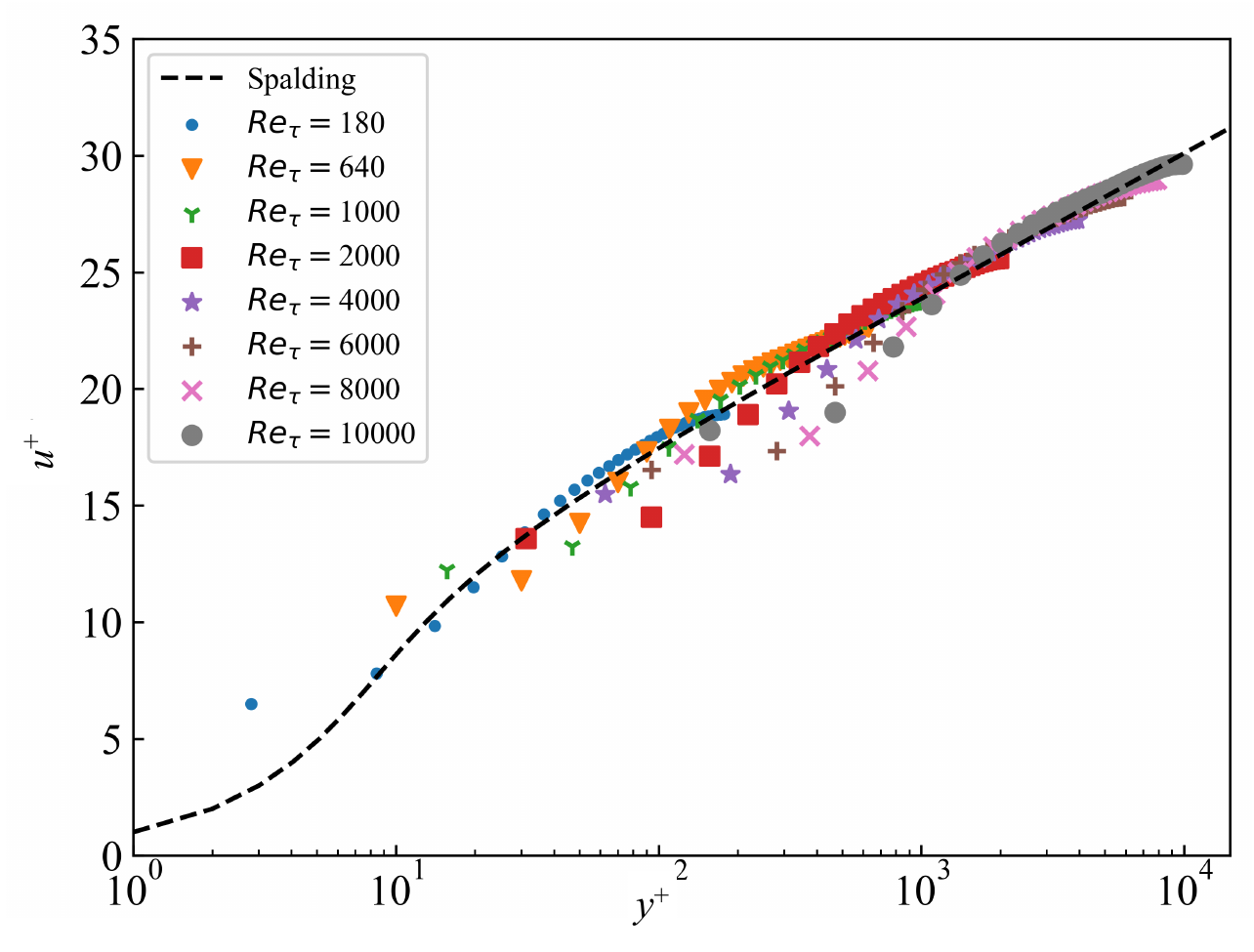}
\caption{Average profiles of streamwise velocity along wall-normal directions for $Re_{\tau}=180-1 \times 10^4$ in wall modeled LES. }
\label{fig:wmles}
 \end{figure}

Figure~\ref{fig:wmles} presents the time-averaged streamwise velocity profiles normalized in wall units. 
The predicted mean velocity profiles match closely with the target Spalding law across all $Re_\tau$, demonstrating that the implemented wall model successfully compensates for the under-resolved near-wall region. As expected, the log-layer mismatch becomes more prominent at lower $Re_\tau$, particularly in the $Re_\tau = 180$ case, which is a known artifact of equilibrium wall models on coarse grids~\cite{larsson2016large}. Nevertheless, the performance improves significantly at higher Reynolds numbers, with excellent agreement observed at $Re_\tau = 2000$ and beyond. These results validate the robustness and scalability of the wall-modeled LES implementation in Diff-FlowFSI. Future enhancements may involve physics-informed extensions~\cite{yang2017log} or integration of trainable wall models using neural networks~\cite{lee2023artificial} within the differentiable programming framework to further reduce log-layer mismatches and improve model adaptability.

\subsection{Computational efficiency and scalability}\label{s:scalability}

We evaluate the computational performance of Diff-FlowFSI in terms of scalability and efficiency by comparing it against OpenFOAM v2312, a widely used open source CFD solver. The benchmarking case is turbulent channel flow at $Re_\tau = 180$, and the reported runtime corresponds to the cost of simulating one flow-through time. All Diff-FlowFSI simulations are executed on a single NVIDIA A100 GPU, while OpenFOAM runs are conducted on a server equipped with AMD EPYC 7643 CPUs~\footnote{Each EPYC 7643 has 48 CPU cores. For OpenFOAM cases with fewer than 48 MPI ranks, we bind execution to a single CPU to avoid non-uniform memory access (NUMA) effects.} and 16-channel 3200 MT/s DDR4 memory (1 TB per socket).  

\begin{figure}[t!]
\centering
\includegraphics[width=1.0\textwidth]{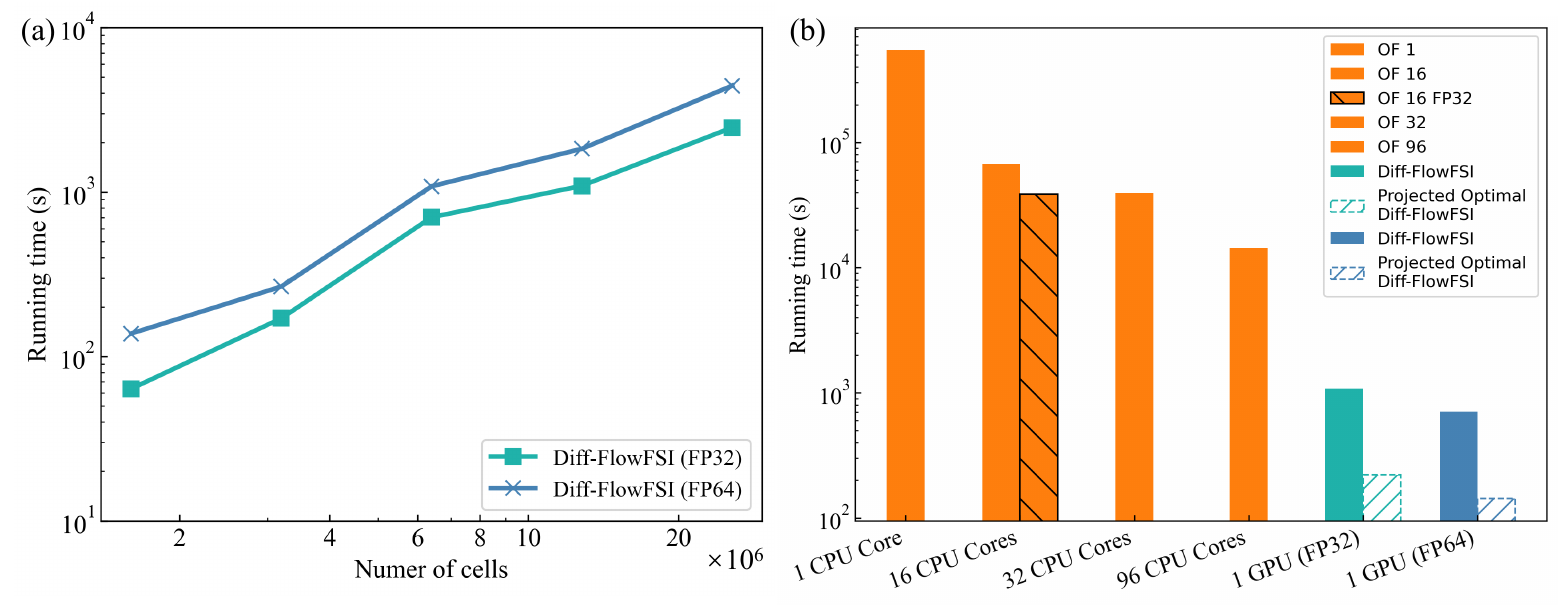}
\caption{ Running time of simulating turbulent channel flow at $Re_\tau = 180$ for one flow-through time: (a) Diff-FlowFSI performance on a single NVIDIA A100 PCIE with varying mesh sizes, and (b) comparison with OpenFOAM running on AMD EPYC 7643 CPU(s) with domain decomposition for a 6-million-cell mesh. The JAX and CUDA versions used are 0.4.19 and CUDA 12, respectively.}
\label{fig:cost}
 \end{figure}

Figure~\ref{fig:cost}(a) presents the runtime scaling of Diff-FlowFSI as a function of grid size under both single precision (FP32) and double precision (FP64) floating-point formats. The solver exhibits near-linear scaling with respect to the number of grid cells, demonstrating its suitability for high-resolution simulations. Importantly, even in FP64 mode, the additional computational cost remains moderate (approximately $1.8 \times$ slower than FP32), validating the practicality of using double precision for high-fidelity scientific computing.

Figure~\ref{fig:cost}(b) compares the performance of Diff-FlowFSI with OpenFOAM for a representative simulation using 6 million cells. In its default configuration, Diff-FlowFSI achieves a dramatic speedup of over $700\times$ compared to OpenFOAM executed on a single CPU core. When domain decomposition is employed in OpenFOAM using 16 cores, Diff-FlowFSI still outpaces it by approximately $90\times$. Although parallelization in OpenFOAM improves wall-clock time, the scalability is sublinear — increasing the core count to 96 only yields a $38\times$ speedup relative to the single-core case. In contrast, Diff-FlowFSI achieves competitive performance on a single GPU without requiring distributed memory or MPI-based parallelization. 
The observed speedups are particularly notable considering that Diff-FlowFSI uses a standard conjugate gradient method for solving the pressure Poisson equation, without preconditioning. In contrast, OpenFOAM uses a multigrid-preconditioned solver that converges in under 10 iterations per step. Despite this advantage, OpenFOAM is still significantly slower, primarily due to its CPU-centric design, reliance on scalar for-loops, and higher memory bandwidth latency. Diff-FlowFSI's performance benefits from full GPU vectorization, JIT compilation, and parallel loop fusion via scan, as discussed in Section~\ref{s:programm}.

Additionally, we assess the impact of floating-point precision on solver performance. While OpenFOAM uses double precision by default, Diff-FlowFSI can operate in both FP32 and FP64 modes. As shown in Figure~\ref{fig:cost}(b), both solvers benefit from switching to FP32, with an observed speedup of approximately $1.7\times$. Even in FP64 mode, Diff-FlowFSI maintains a performance advantage exceeding $60\times$ compared to OpenFOAM with 16 CPU cores.

These results demonstrate that Diff-FlowFSI offers a highly efficient, GPU-native architecture for high-fidelity simulations. Its single-GPU speed rivals that of large-scale CPU clusters, eliminating the need for domain decomposition in many applications. Looking forward, Diff-FlowFSI supports modular extension to multi-GPU parallelism and distributed training loops, opening new possibilities for integrating real-time simulation and machine learning workflows within a unified differentiable framework. This capability is particularly advantageous for hybrid physics-ML modeling, where simulation data and model training can proceed concurrently at scale.

\section{Inverse modeling and hybrid learning applications} \label{s:ad}

This section demonstrates the differentiable programming capability of Diff-FlowFSI for solving inverse problems and training hybrid neural models. Three applications are presented: parameter inference from sparse observations, hybrid neural differentiable models for accelerated FSI simulations, and hybrid neural solver architecture design for homogeneous isotropic turbulence.  

\subsection{Physical parameter inversion from sparse observations} \label{s:predicting}

To validate the AD capability of Diff-FlowFSI, we first consider an inverse problem: inferring unknown physical parameters from sparse measurements. In practice, structural responses such as displacement and acceleration are commonly measured by sensors, and these data can be leveraged to infer unobserved physical parameters, such as material properties. As a proof-of-concept, we aim to recover the unknown spring stiffness in a 2DOF VIV system using time-resolved displacement observations. 

\begin{figure}[t!]
\centering
\includegraphics[width=0.8\textwidth]{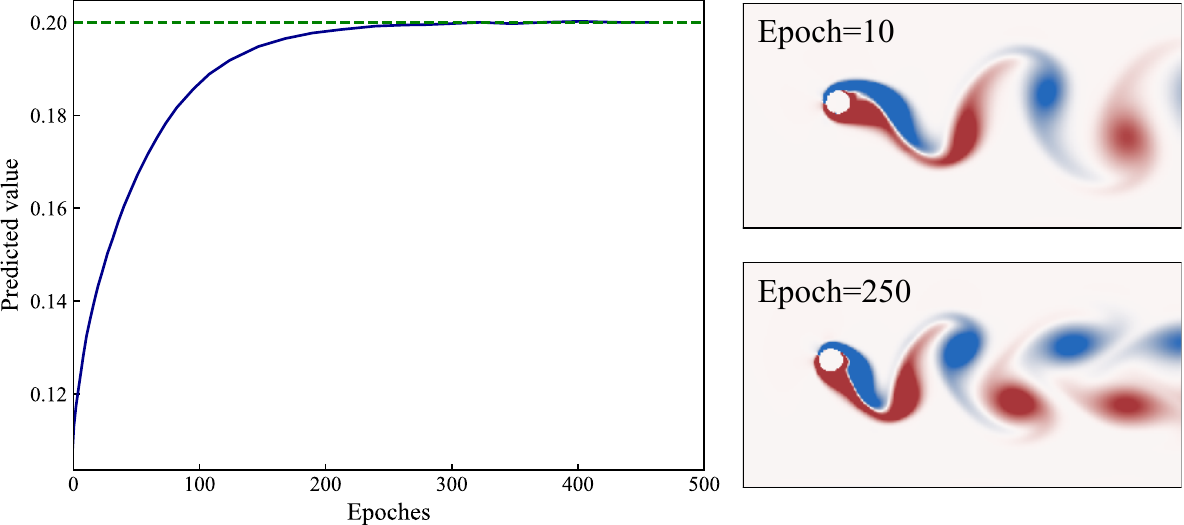}
\caption{Inverse modeling using AD in Diff-FlowFSI: prediction of unknown spring stiffness and reconstruction of high-dimensional flow field based on sparse structural displacements.}
\label{fig:reverse-VIV}
 \end{figure}

The ground-truth spring stiffness is set as 0.2 and the corresponding displacement are obtained from a forward simulation described in Sec.~\ref{s: 2dof}, which are used as observation data. The spring stiffness is treated as a learnable parameter, initialized with a rough guess (0.1), and optimized using gradient descent with a learning rate of $1\times10^{-3}$. The objective is to minimize the mismatch between predicted and observed displacements using backpropagation through the solver. As shown in Fig.\ref{fig:reverse-VIV}, Diff-FlowFSI successfully recovers the true stiffness after 250 learning epochs.

This example highlights the capability of Diff-FlowFSI in solving inverse problems through gradient-based optimization. More notably, the framework enables the recovery of high-dimensional latent flow fields (e.g., velocity) from low-dimensional scalar signals (e.g., displacement). This result suggests the potential for sensor-driven flow field inference as a computational alternative to data-intensive methods like PIV, particularly in settings where experimental measurements are sparse or incomplete.

\subsection{Hybrid learning with physics-integrated neural differentiable models}

\subsubsection{Hybrid neural solver for accelerated FSI simulations}

\begin{figure}[hpt!]
\centering
\includegraphics[width=0.95\textwidth]{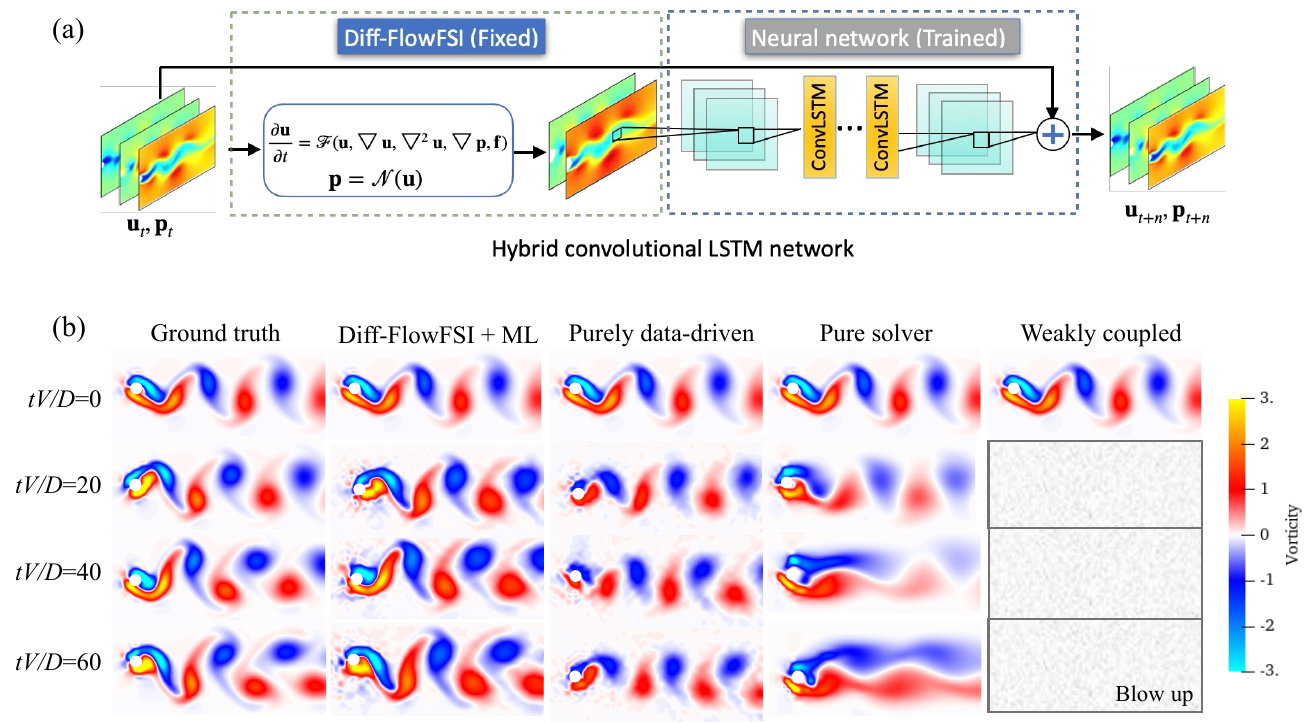}
\caption{Hybrid neural Diff-FlowFSI solver for predicting VIV dynamics of rigid and flexible structures: (a) the schematic showing the hybrid model integrates a neural correction term into the differentiable physics solver; (b) the comparison of vorticity for the rigid cylinder in unseen stiffness.}
\label{fig:VIV_ML}
 \end{figure}

We proposed a hybrid differentiable learning model within the Diff-FlowFSI framework to accelerate FSI simulations, as detailed in~\cite{fan2023differentiable}. The central idea is to integrate a coarse-resolution physical solver with data-driven correction terms within a unified, differentiable architecture. This hybrid model is implemented as a convolutional LSTM network, where part of the convolutional layers are non-trainable and correspond to the built-in numerical operators from Diff-FlowFSI, as illustrated in Figure~\ref{fig:VIV_ML}(a). These fixed operators enforce the governing physics, while the trainable neural components learn to correct the coarse solutions and enhance predictive accuracy. The hybrid prediction model can be conceptually expressed as: 
\begin{equation}
    \mathbf{u}_{t+1}=\mathcal{F}(\mathbf{x}, \mathbf{u}_t, \lambda) + \mathcal{N}(\mathbf{x}, \mathbf{u}_t, \lambda;\boldsymbol{ \theta}).
    \label{eq:pde-neural-corr}
\end{equation}
where $\mathcal{F}$ denotes the discretized equations implemented in Diff-FlowFSI, $\mathcal{N}$ is the learnable neural correction network, $\mathbf{u}$ represents the predicted state variables, $\mathbf{x}$ is the spatial coordinate, $\lambda$ denotes relevant physical parameters, and $\boldsymbol{\theta}$ are trainable weights.

The fully differentiable nature of Diff-FlowFSI allows the hybrid model to be trained end-to-end using \textit{a posteriori} loss functions defined on physical quantities of interest, such as structural displacement or hydrodynamic forces. By embedding the governing equations as part of the sequence model's internal layers, the hybrid architecture mitigates error accumulation and improves robustness, addressing two of the major limitations of traditional black-box forecasting models.
We validate this framework on two representative benchmarks involving VIV of rigid and flexible cylinders. As shown in Figure~\ref{fig:VIV_ML}(b), the hybrid model accurately predicts long-time structural displacements and fluid forces, while maintaining physical consistency. Compared to black-box neural surrogates or weakly coupled hybrid neural model (without differentiable co-training), the hybrid model achieves superior generalization and stability, particularly in long rollouts and unseen parameter regimes. The differentiable architecture also facilitates efficient training and optimization, making the framework suitable for data-limited, high-fidelity scientific computing applications.

\subsubsection{Hybrid neural model architecture design for homogeneous isotropic turbulence}

We conducted a detailed study on hybrid neural solver architectures within the Diff-FlowFSI framework, focusing on the modeling of homogeneous isotropic turbulence (HIT)~\cite{fan2025neural}. In this context, we explore two distinct integration strategies for incorporating neural networks into PDE solvers within $\partial$P framework. The first strategy appends trainable neural correction terms to the governing equations, whereas the second adopts a deeper fusion approach, wherein the neural network learns the flux interpolation schemes within the PDE solver itself.

\begin{figure}[t!]
\centering
\includegraphics[width=0.95\textwidth]{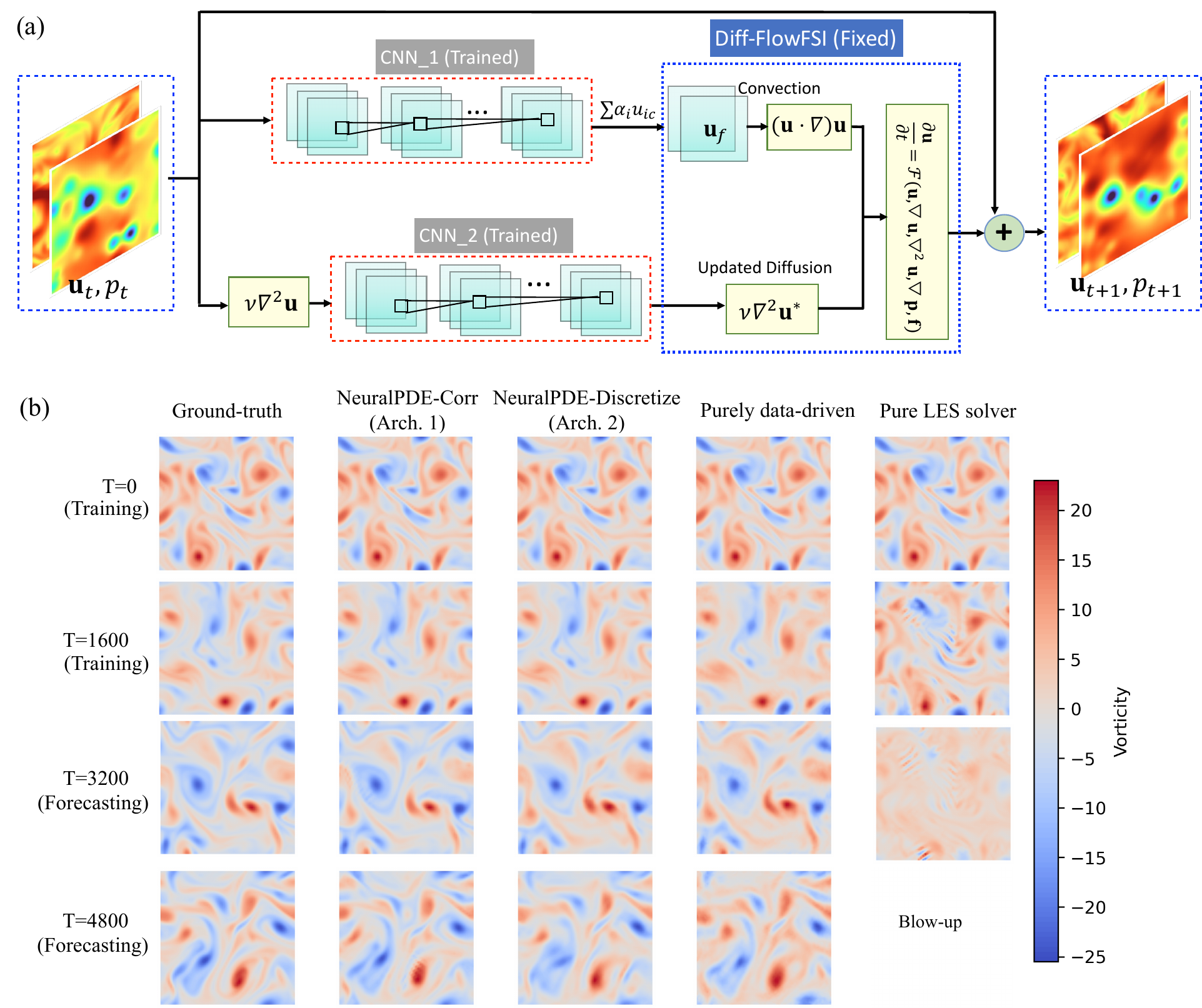}
\caption{Hybrid neural solver architecture design for turbulence: (a) schematic of deep fusion architecture integrating ML into the numerical pipeline of Diff-FlowFSI; (b) comparison of turbulence rollouts using different integration architectures. Adapted from~\cite{fan2025neural}.}
\label{fig:ML_HIT_contour}
 \end{figure}

The first architecture—resembling Eq.~\ref{eq:pde-neural-corr}, introduced earlier, augments the existing physical solver by learning supplementary closure terms such as subgrid-scale (SGS) stresses or Reynolds-averaged turbulence models in a multi-resolution manner. This modular approach treats the neural network as a post-processing correction, making it robust and straightforward to train. In contrast, the second architecture is expressed as:
\begin{equation}
\mathbf{u}_{t+1}=\mathcal{F} ( \mathbf{x}, \mathbf{u}_t, \lambda , \mathcal{N}(\mathbf{x}, \mathbf{u}_t, \lambda;\boldsymbol{ \theta}) ),
\label{eq:pde-neural-num}
\end{equation}
where the neural network $\mathcal{N}$ replaces selected numerical components (e.g., flux interpolation or viscous term approximations) within the core PDE solver $\mathcal{F}$. This deeper integration, illustrated in Figure~\ref{fig:ML_HIT_contour}(a), enables the hybrid model to learn implicit numerical schemes directly from data, offering potentially higher fidelity but requiring significantly more careful design.

Figure~\ref{fig:ML_HIT_contour}(b) compares the rollout performance of these architectures for HIT prediction tasks. Both models successfully capture the large-scale turbulence evolution over long horizons. However, the deeper fusion model is more sensitive to numerical instability. Minor perturbations introduced by the neural network, especially when directly altering core stencil operations, can degrade the conditioning of the solver and destabilize the rollout. To mitigate this issue, we recommend incorporating stabilization techniques, such as constraining the neural outputs via spectral or physical bounds~\cite{zhuang2021learned}, using normalized activation layers, or embedding energy-preserving inductive biases in the architecture design. These modifications are essential to preserve stability when deploying neural solvers in tightly coupled, differentiable physics simulations.

These findings emphasize the importance of carefully balancing physical fidelity and architectural complexity in the hybrid ML-solver design. While modular correction terms provide ease of training and strong generalization, deeply integrated schemes promise higher expressiveness at the cost of numerical robustness. The Diff-FlowFSI framework supports both, offering a versatile platform for exploring hybrid physics–ML models in flow/FSI problems.

\section{Conclusion}\label{s:conclusion}

We have introduced Diff-FlowFSI, a fully differentiable, GPU-accelerated computational platform for simulating unsteady turbulent flow and fluid–structure interactions with end-to-end gradient access. Implemented in JAX and structured around the immersed boundary method and finite volume discretization, Diff-FlowFSI provides a unified framework for high-fidelity forward simulations, inverse modeling, and physics-informed machine learning. The solver supports both 2D and 3D domains, accommodates static and dynamic solid bodies (rigid or flexible), and enables strong two-way coupling between fluid and structural solvers. Its extensible architecture facilitates arbitrary geometry definitions and automatic differentiation throughout the simulation pipeline.

We validated Diff-FlowFSI across a comprehensive suite of canonical benchmarks—including vortex shedding, vortex-induced vibration, flexible plate deformation, wall-bounded turbulence, and separated flows, demonstrating accurate predictions of physical quantities such as pressure, forces, vortex dynamics, and turbulent statistics. This fully vectorized solver exhibits excellent computational scalability on modern GPUs and outperforms traditional CPU-based solvers like OpenFOAM by up to two orders of magnitude in runtime efficiency. Notably, it supports both single and double precision computations while maintaining high fidelity. We further demonstrated the utility of automatic differentiation in two key applications: (i) inverse modeling via gradient-based recovery of unknown parameters from sparse indirect measurements, and (ii) hybrid neural modeling by integrating deep learning with differentiable physics solvers for long-time forecasting of FSI and turbulence. These applications underscore Diff-FlowFSI's potential as a scientific machine learning platform that tightly couples data, physics, and optimization.

Looking forward, several avenues for future development remain:
\begin{enumerate}
    \item Enhanced numerical fidelity: Incorporating advanced discretization schemes, such as pressure-implicit methods and dynamic Smagorinsky SGS models, would broaden the scope and accuracy of high-Reynolds-number simulations.
    \item Full 3D strong coupling and interface accuracy: Extending the current 2D strong coupling framework to three dimensions, along with improved Lagrangian–Eulerian interpolation strategies, will be critical for simulating complex structural deformations and achieving second-order accuracy near interfaces.
    \item Scalable parallelism: Although Diff-FlowFSI supports GPU acceleration and limited data parallelism, future work will focus on domain decomposition and distributed solver backends to enable multi-GPU and multi-node scalability for very large-scale simulations.
\end{enumerate}

Through its differentiable, modular, and high-performance design, Diff-FlowFSI bridges the gap between conventional CFD solvers and emerging differentiable programming paradigms. It offers a principled and extensible foundation for future advances in inverse design, uncertainty quantification, data assimilation, and hybrid AI–physics modeling. 

\section*{Acknowledgements}
 The authors would like to acknowledge the funds from Office of Naval Research under award numbers N00014-23-1-2071 and National Science Foundation under award numbers OAC-2047127. XF would also like to acknowledge the fellowship provided by the Environmental Change Initiative and Center for Sustainable Energy at University of Notre Dame. Thanks for the discussion from Akshay Thakur for the simulation setup in 3D channel flow.

\section*{Compliance with Ethical Standards}
Conflict of Interest: The authors declare that they have no conflict of interest.


\bibliographystyle{elsarticle-num}
\bibliography{ref}

\begin{thebibliography}{10}
\expandafter\ifx\csname url\endcsname\relax
  \def\url#1{\texttt{#1}}\fi
\expandafter\ifx\csname urlprefix\endcsname\relax\def\urlprefix{URL }\fi
\expandafter\ifx\csname href\endcsname\relax
  \def\href#1#2{#2} \def\path#1{#1}\fi

\bibitem{oberkampf2002verification}
W.~L. Oberkampf, T.~G. Trucano, Verification and validation in computational
  fluid dynamics, Progress in aerospace sciences 38~(3) (2002) 209--272.

\bibitem{bhatti2020recent}
M.~M. Bhatti, M.~Marin, A.~Zeeshan, S.~I. Abdelsalam, Recent trends in
  computational fluid dynamics, Frontiers in Physics 8 (2020) 593111.

\bibitem{zawawi2018review}
M.~H. Zawawi, A.~Saleha, A.~Salwa, N.~Hassan, N.~M. Zahari, M.~Z. Ramli, Z.~C.
  Muda, A review: Fundamentals of computational fluid dynamics (cfd), in: AIP
  conference proceedings, Vol. 2030, AIP Publishing, 2018.

\bibitem{ma2015using}
M.~Ma, J.~Lu, G.~Tryggvason, Using statistical learning to close two-fluid
  multiphase flow equations for a simple bubbly system, Physics of Fluids
  27~(9) (2015).

\bibitem{fairbanks2020bi}
H.~R. Fairbanks, L.~Jofre, G.~Geraci, G.~Iaccarino, A.~Doostan, Bi-fidelity
  approximation for uncertainty quantification and sensitivity analysis of
  irradiated particle-laden turbulence, Journal of Computational Physics 402
  (2020) 108996.

\bibitem{sharma2024amrex}
S.~Sharma, R.~Bielawski, O.~Gibson, S.~Zhang, V.~Sharma, A.~H. Rauch, J.~Singh,
  S.~Abisleiman, M.~Ullman, S.~Barwey, et~al., An amrex-based compressible
  reacting flow solver for high-speed reacting flows relevant to hypersonic
  propulsion, arXiv preprint arXiv:2412.00900 (2024).

\bibitem{fuhg2024review}
J.~N. Fuhg, G.~Anantha~Padmanabha, N.~Bouklas, B.~Bahmani, W.~Sun, N.~N.
  Vlassis, M.~Flaschel, P.~Carrara, L.~De~Lorenzis, A review on data-driven
  constitutive laws for solids, Archives of Computational Methods in
  Engineering (2024) 1--43.

\bibitem{fan2021impacts}
X.~Fan, M.~Ge, W.~Tan, Q.~Li, Impacts of coexisting buildings and trees on the
  performance of rooftop wind turbines: An idealized numerical study, Renewable
  Energy 177 (2021) 164--180.

\bibitem{xiao2016quantifying}
H.~Xiao, J.-L. Wu, J.-X. Wang, R.~Sun, C.~Roy, Quantifying and reducing
  model-form uncertainties in reynolds-averaged navier--stokes simulations: A
  data-driven, physics-informed bayesian approach, Journal of Computational
  Physics 324 (2016) 115--136.

\bibitem{duraisamy2019turbulence}
K.~Duraisamy, G.~Iaccarino, H.~Xiao, Turbulence modeling in the age of data,
  Annual review of fluid mechanics 51~(1) (2019) 357--377.

\bibitem{jofre2022rapid}
L.~Jofre, A.~Doostan, Rapid aerodynamic shape optimization under uncertainty
  using a stochastic gradient approach, Structural and Multidisciplinary
  Optimization 65~(7) (2022) 196.

\bibitem{brunton2020machine}
S.~L. Brunton, B.~R. Noack, P.~Koumoutsakos, Machine learning for fluid
  mechanics, Annual review of fluid mechanics 52~(1) (2020) 477--508.

\bibitem{du2022deep}
P.~Du, X.~Zhu, J.-X. Wang, Deep learning-based surrogate model for
  three-dimensional patient-specific computational fluid dynamics, Physics of
  Fluids 34~(8) (2022) 081906.

\bibitem{han2022predicting}
X.~Han, H.~Gao, T.~Pfaff, J.-X. Wang, L.~Liu,
  \href{https://openreview.net/forum?id=XctLdNfCmP}{Predicting physics in
  mesh-reduced space with temporal attention}, in: International Conference on
  Learning Representations, 2022.
\newline\urlprefix\url{https://openreview.net/forum?id=XctLdNfCmP}

\bibitem{du2024confild}
P.~Du, M.~H. Parikh, X.~Fan, X.-Y. Liu, J.-X. Wang, Conditional neural field
  latent diffusion model for generating spatiotemporal turbulence, Nature
  Communications (2024).

\bibitem{karniadakis2021physics}
G.~E. Karniadakis, I.~G. Kevrekidis, L.~Lu, P.~Perdikaris, S.~Wang, L.~Yang,
  Physics-informed machine learning, Nature Reviews Physics 3~(6) (2021)
  422--440.

\bibitem{raissi2019physics}
M.~Raissi, P.~Perdikaris, G.~E. Karniadakis, Physics-informed neural networks:
  A deep learning framework for solving forward and inverse problems involving
  nonlinear partial differential equations, Journal of Computational physics
  378 (2019) 686--707.

\bibitem{sun2020surrogate}
L.~Sun, H.~Gao, S.~Pan, J.-X. Wang, Surrogate modeling for fluid flows based on
  physics-constrained deep learning without simulation data, Computer Methods
  in Applied Mechanics and Engineering 361 (2020) 112732.

\bibitem{arzani2021uncovering}
A.~Arzani, J.-X. Wang, R.~M. D'Souza, Uncovering near-wall blood flow from
  sparse data with physics-informed neural networks, Physics of Fluids 33~(7)
  (2021) 071905.
\newblock \href {https://doi.org/10.1063/5.0055600}
  {\path{doi:10.1063/5.0055600}}.

\bibitem{chen2020physics}
Y.~Chen, L.~Lu, G.~E. Karniadakis, L.~Dal~Negro, Physics-informed neural
  networks for inverse problems in nano-optics and metamaterials, Optics
  express 28~(8) (2020) 11618--11633.

\bibitem{gao2021phygeonet}
H.~Gao, L.~Sun, J.-X. Wang, {PhyGeoNet:} physics-informed geometry-adaptive
  convolutional neural networks for solving parameterized steady-state {PDEs}
  on irregular domain, Journal of Computational Physics 428 (2021) 110079.

\bibitem{li2025physics}
R.~Li, J.~Zhou, J.-X. Wang, T.~Luo, Physics-informed bayesian neural networks
  for solving phonon boltzmann transport equation in forward and inverse
  problems with sparse and noisy data, ASME Journal of Heat and Mass Transfer
  147~(3) (2025) 032501.

\bibitem{chen2021physics}
Z.~Chen, Y.~Liu, H.~Sun, Physics-informed learning of governing equations from
  scarce data, Nature communications 12~(1) (2021) 6136.

\bibitem{krishnapriyan2021characterizing}
A.~Krishnapriyan, A.~Gholami, S.~Zhe, R.~Kirby, M.~W. Mahoney, Characterizing
  possible failure modes in physics-informed neural networks, Advances in
  Neural Information Processing Systems 34 (2021) 26548--26560.

\bibitem{wang2022and}
S.~Wang, X.~Yu, P.~Perdikaris, When and why pinns fail to train: A neural
  tangent kernel perspective, Journal of Computational Physics 449 (2022)
  110768.

\bibitem{wang2021understanding}
S.~Wang, Y.~Teng, P.~Perdikaris, Understanding and mitigating gradient flow
  pathologies in physics-informed neural networks, SIAM Journal on Scientific
  Computing 43~(5) (2021) A3055--A3081.

\bibitem{brenner2019perspective}
M.~Brenner, J.~Eldredge, J.~Freund, Perspective on machine learning for
  advancing fluid mechanics, Physical Review Fluids 4~(10) (2019) 100501.

\bibitem{wang2017physics}
J.-X. Wang, J.-L. Wu, H.~Xiao, Physics-informed machine learning approach for
  reconstructing reynolds stress modeling discrepancies based on {DNS} data,
  Physical Review Fluids 2~(3) (2017) 034603.

\bibitem{tompson2017accelerating}
J.~Tompson, K.~Schlachter, P.~Sprechmann, K.~Perlin, Accelerating eulerian
  fluid simulation with convolutional networks, in: International conference on
  machine learning, PMLR, 2017, pp. 3424--3433.

\bibitem{wu2019reynolds}
J.~Wu, H.~Xiao, R.~Sun, Q.~Wang, Reynolds-averaged navier--stokes equations
  with explicit data-driven reynolds stress closure can be ill-conditioned,
  Journal of Fluid Mechanics 869 (2019) 553--586.

\bibitem{taghizadeh2020turbulence}
S.~Taghizadeh, F.~D. Witherden, S.~S. Girimaji, Turbulence closure modeling
  with data-driven techniques: physical compatibility and consistency
  considerations, New Journal of Physics 22~(9) (2020) 093023.

\bibitem{fan2023differentiable}
X.~Fan, J.-X. Wang, Differentiable hybrid neural modeling for fluid-structure
  interaction, arXiv preprint arXiv:2303.12971 (2023).

\bibitem{liu2024multi}
X.-Y. Liu, M.~Zhu, L.~Lu, H.~Sun, J.-X. Wang, Multi-resolution partial
  differential equations preserved learning framework for spatiotemporal
  dynamics, Communications Physics 7~(1) (2024) 31.

\bibitem{fan2025neural}
X.~Fan, D.~Akhare, J.-X. Wang, Neural differentiable modeling with
  diffusion-based super-resolution for two-dimensional spatiotemporal
  turbulence, Computer Methods in Applied Mechanics and Engineering 433 (2025)
  117478.

\bibitem{kochkov2021machine}
D.~Kochkov, J.~A. Smith, A.~Alieva, Q.~Wang, M.~P. Brenner, S.~Hoyer, Machine
  learning--accelerated computational fluid dynamics, Proceedings of the
  National Academy of Sciences 118~(21) (2021) e2101784118.

\bibitem{akhare2023diffhybrid}
D.~Akhare, T.~Luo, J.-X. Wang, Diffhybrid-uq: Uncertainty quantification for
  differentiable hybrid neural modeling, arXiv preprint arXiv:2401.00161
  (2023).

\bibitem{akhare2023physics}
D.~Akhare, T.~Luo, J.-X. Wang, Physics-integrated neural differentiable
  {(PiNDiff)} model for composites manufacturing, Computer Methods in Applied
  Mechanics and Engineering 406 (2023) 115902.

\bibitem{akhare2024probabilistic}
D.~Akhare, Z.~Chen, R.~Gulotty, T.~Luo, J.-X. Wang, Probabilistic
  physics-integrated neural differentiable modeling for isothermal chemical
  vapor infiltration process, npj Computational Materials 10~(1) (2024) 120.

\bibitem{list2022learned}
B.~List, L.-W. Chen, N.~Thuerey, Learned turbulence modelling with
  differentiable fluid solvers: physics-based loss functions and optimisation
  horizons, Journal of Fluid Mechanics 949 (2022) A25.

\bibitem{belbute2020combining}
F.~D.~A. Belbute-Peres, T.~Economon, Z.~Kolter, Combining differentiable pde
  solvers and graph neural networks for fluid flow prediction, in:
  international conference on machine learning, PMLR, 2020, pp. 2402--2411.

\bibitem{shankar2025differentiable}
V.~Shankar, D.~Chakraborty, V.~Viswanathan, R.~Maulik, Differentiable
  turbulence: Closure as a partial differential equation constrained
  optimization, Physical Review Fluids 10~(2) (2025) 024605.

\bibitem{shang2025jax}
W.~Shang, J.~Zhou, J.~Panda, Z.~Xu, Y.~Liu, P.~Du, J.-X. Wang, T.~Luo, Jax-bte:
  A gpu-accelerated differentiable solver for phonon boltzmann transport
  equations, arXiv preprint arXiv:2503.23657 (2025).

\bibitem{bischof2008implementation}
C.~H. Bischof, P.~D. Hovland, B.~Norris, On the implementation of automatic
  differentiation tools, Higher-Order and Symbolic Computation 21 (2008)
  311--331.

\bibitem{mcnamara2004fluid}
A.~McNamara, A.~Treuille, Z.~Popovi{\'c}, J.~Stam, Fluid control using the
  adjoint method, ACM Transactions On Graphics (TOG) 23~(3) (2004) 449--456.

\bibitem{hinterberger2010automatic}
C.~Hinterberger, M.~Olesen, Automatic geometry optimization of exhaust systems
  based on sensitivities computed by a continuous adjoint cfd method in
  openfoam, Tech. rep., SAE Technical Paper (2010).

\bibitem{shi2020natural}
Y.~Shi, C.~A. Mader, S.~He, G.~L. Halila, J.~R. Martins, Natural laminar-flow
  airfoil optimization design using a discrete adjoint approach, AIAA Journal
  58~(11) (2020) 4702--4722.

\bibitem{mader2008adjoint}
C.~A. Mader, J.~R. Martins, J.~J. Alonso, E.~Van Der~Weide, Adjoint: An
  approach for the rapid development of discrete adjoint solvers, AIAA journal
  46~(4) (2008) 863--873.

\bibitem{holl2020phiflow}
P.~Holl, V.~Koltun, K.~Um, N.~Thuerey, phiflow: A differentiable pde solving
  framework for deep learning via physical simulations, in: NeurIPS workshop,
  Vol.~2, 2020.

\bibitem{RN1108}
N.~A.~A. Deniz A.~Bezgin, Aaron B.~Buhendwa, Jax-fluids: A fully-differentiable
  high-order computational fluid dynamics solver for compressible two-phase
  flows, arXiv preprint arXiv (2022).

\bibitem{boustani2021immersed}
J.~Boustani, M.~F. Barad, C.~C. Kiris, C.~Brehm, An immersed boundary
  fluid--structure interaction method for thin, highly compliant shell
  structures, Journal of Computational Physics 438 (2021) 110369.

\bibitem{ramaswamy1987arbitrary}
B.~Ramaswamy, M.~Kawahara, Arbitrary lagrangian--eulerianc finite element
  method for unsteady, convective, incompressible viscous free surface fluid
  flow, International Journal for Numerical Methods in Fluids 7~(10) (1987)
  1053--1075.

\bibitem{malecha2011gpu}
Z.~Malecha, {\L}.~Miros{\l}aw, T.~Tomczak, Z.~Koza, M.~Matyka, W.~Tarnawski,
  D.~Szczerba, et~al., Gpu-based simulation of 3d blood flow in abdominal aorta
  using openfoam, Archives of Mechanics 63~(2) (2011) 137--161.

\bibitem{rathnayake2017openfoam}
T.~Rathnayake, S.~Jayasena, M.~Narayana, Openfoam on gpus using amgx, in:
  Proceedings of the 25th High Performance Computing Symposium, 2017, pp.
  1--12.

\bibitem{harris2020array}
C.~R. Harris, K.~J. Millman, S.~J. Van Der~Walt, R.~Gommers, P.~Virtanen,
  D.~Cournapeau, E.~Wieser, J.~Taylor, S.~Berg, N.~J. Smith, et~al., Array
  programming with numpy, Nature 585~(7825) (2020) 357--362.

\bibitem{peskin2002immersed}
C.~S. Peskin, The immersed boundary method, Acta numerica 11 (2002) 479--517.

\bibitem{RN1208}
O.~A. Bauchau, J.~I. Craig, Euler-Bernoulli beam theory, Springer, 2009, pp.
  173--221.

\bibitem{peskin1972flow}
C.~S. Peskin, Flow patterns around heart valves: a numerical method, Journal of
  computational physics 10~(2) (1972) 252--271.

\bibitem{uhlmann2005immersed}
M.~Uhlmann, An immersed boundary method with direct forcing for the simulation
  of particulate flows, Journal of computational physics 209~(2) (2005)
  448--476.

\bibitem{zienkiewicz2005finite}
O.~C. Zienkiewicz, R.~L. Taylor, The finite element method for solid and
  structural mechanics, Elsevier, 2005.

\bibitem{liu1995formulation}
M.~Liu, D.~G. Gorman, Formulation of rayleigh damping and its extensions,
  Computers \& structures 57~(2) (1995) 277--285.

\bibitem{ji2012novel}
C.~Ji, A.~Munjiza, J.~Williams, A novel iterative direct-forcing immersed
  boundary method and its finite volume applications, Journal of Computational
  Physics 231~(4) (2012) 1797--1821.

\bibitem{baydin2018automatic}
A.~G. Baydin, B.~A. Pearlmutter, A.~A. Radul, J.~M. Siskind, Automatic
  differentiation in machine learning: a survey, Journal of Marchine Learning
  Research 18 (2018) 1--43.

\bibitem{akhare2025implicit}
D.~Akhare, P.~Du, T.~Luo, J.-X. Wang, Implicit neural differential model for
  spatiotemporal dynamics, arXiv preprint arXiv:2504.02260 (2025).

\bibitem{blondel2024elements-88d}
M.~Blondel, V.~Roulet, The elements of differentiable programming
  {arXiv}:2403.14606v1 [cs.{LG}] 21 mar 2024 (2024).

\bibitem{jaxopt_implicit_diff}
M.~Blondel, Q.~Berthet, M.~Cuturi, R.~Frostig, S.~Hoyer, F.~Llinares-L{\'o}pez,
  F.~Pedregosa, J.-P. Vert, Efficient and modular implicit differentiation,
  arXiv preprint arXiv:2105.15183 (2021).

\bibitem{henderson1997nonlinear}
R.~D. Henderson, Nonlinear dynamics and pattern formation in turbulent wake
  transition, Journal of fluid mechanics 352 (1997) 65--112.

\bibitem{wieselsberger1921neuere}
C.~v. Wieselsberger, Neuere feststellungen under die gesetze des flussigkeits
  und luftwiderstandes, Phys. z. 22 (1921) 321--328.

\bibitem{posdziech2007systematic}
O.~Posdziech, R.~Grundmann, A systematic approach to the numerical calculation
  of fundamental quantities of the two-dimensional flow over a circular
  cylinder, Journal of fluids and structures 23~(3) (2007) 479--499.

\bibitem{park1998numerical}
J.~Park, K.~Kwon, H.~Choi, Numerical solutions of flow past a circular cylinder
  at reynolds numbers up to 160, KSME international Journal 12 (1998)
  1200--1205.

\bibitem{williamson1989oblique}
C.~H. Williamson, Oblique and parallel modes of vortex shedding in the wake of
  a circular cylinder at low reynolds numbers, Journal of Fluid Mechanics 206
  (1989) 579--627.

\bibitem{chen2018vortex}
W.~Chen, C.~Ji, J.~Williams, D.~Xu, L.~Yang, Y.~Cui, Vortex-induced vibrations
  of three tandem cylinders in laminar cross-flow: Vibration response and
  galloping mechanism, Journal of Fluids and Structures 78 (2018) 215--238.

\bibitem{bao2012two}
Y.~Bao, C.~Huang, D.~Zhou, J.~Tu, Z.~Han, Two-degree-of-freedom flow-induced
  vibrations on isolated and tandem cylinders with varying natural frequency
  ratios, Journal of Fluids and Structures 35 (2012) 50--75.

\bibitem{gsell2021direct}
S.~Gsell, J.~Favier, Direct-forcing immersed-boundary method: A simple
  correction preventing boundary slip error, Journal of Computational Physics
  435 (2021) 110265.

\bibitem{gabbai2005overview}
R.~D. Gabbai, H.~Benaroya, An overview of modeling and experiments of
  vortex-induced vibration of circular cylinders, Journal of sound and
  vibration 282~(3-5) (2005) 575--616.

\bibitem{srinil2013two}
N.~Srinil, H.~Zanganeh, A.~Day, Two-degree-of-freedom viv of circular cylinder
  with variable natural frequency ratio: Experimental and numerical
  investigations, Ocean Engineering 73 (2013) 179--194.

\bibitem{prasanth2008vortex}
T.~Prasanth, S.~Mittal, Vortex-induced vibrations of a circular cylinder at low
  reynolds numbers, Journal of Fluid Mechanics 594 (2008) 463--491.

\bibitem{zhang2020fluid}
X.~Zhang, G.~He, X.~Zhang, Fluid--structure interactions of single and dual
  wall-mounted 2d flexible filaments in a laminar boundary layer, Journal of
  Fluids and Structures 92 (2020) 102787.

\bibitem{pfister2020fluid}
J.-L. Pfister, O.~Marquet, Fluid--structure stability analyses and nonlinear
  dynamics of flexible splitter plates interacting with a circular cylinder
  flow, Journal of Fluid Mechanics 896 (2020) A24.

\bibitem{moser1999direct}
R.~D. Moser, J.~Kim, N.~N. Mansour, Direct numerical simulation of turbulent
  channel flow up to re $\tau$= 590, Physics of fluids 11~(4) (1999) 943--945.

\bibitem{kim1987turbulence}
J.~Kim, P.~Moin, R.~Moser, Turbulence statistics in fully developed channel
  flow at low reynolds number, Journal of fluid mechanics 177 (1987) 133--166.

\bibitem{abe2001direct}
H.~Abe, H.~Kawamura, Y.~Matsuo, Direct numerical simulation of a fully
  developed turbulent channel flow with respect to the reynolds number
  dependence, J. Fluids Eng. 123~(2) (2001) 382--393.

\bibitem{rai1991direct}
M.~M. Rai, P.~Moin, Direct simulations of turbulent flow using
  finite-difference schemes, Journal of computational physics 96~(1) (1991)
  15--53.

\bibitem{dean1978reynolds}
R.~B. Dean, Reynolds number dependence of skin friction and other bulk flow
  variables in two-dimensional rectangular duct flow (1978).

\bibitem{steen2017saph}
P.~Steen, W.~Brutsaert, Saph and schoder and the friction law of blasius,
  Annual Review of Fluid Mechanics 49 (2017) 575--582.

\bibitem{santiago2007cfd}
J.~L. Santiago, A.~Martilli, F.~Mart{\'\i}n, Cfd simulation of airflow over a
  regular array of cubes. part i: Three-dimensional simulation of the flow and
  validation with wind-tunnel measurements, Boundary-layer meteorology 122
  (2007) 609--634.

\bibitem{cai2017moving}
S.-G. Cai, A.~Ouahsine, J.~Favier, Y.~Hoarau, Moving immersed boundary method,
  International Journal for Numerical Methods in Fluids 85~(5) (2017) 288--323.

\bibitem{zhou2021analysis}
K.~Zhou, S.~Balachandar, An analysis of the spatio-temporal resolution of the
  immersed boundary method with direct forcing, Journal of Computational
  Physics 424 (2021) 109862.

\bibitem{kim2010simulating}
Y.~Kim, M.-C. Lai, Simulating the dynamics of inextensible vesicles by the
  penalty immersed boundary method, Journal of Computational Physics 229~(12)
  (2010) 4840--4853.

\bibitem{breuer2009flow}
M.~Breuer, N.~Peller, C.~Rapp, M.~Manhart, Flow over periodic hills--numerical
  and experimental study in a wide range of reynolds numbers, Computers \&
  Fluids 38~(2) (2009) 433--457.

\bibitem{douglas2022strouhal}
A.~Douglas, et~al., Strouhal number for vortex-induced vibration excitation of
  long slender structures (2022).

\bibitem{scotti1993generalized}
A.~Scotti, C.~Meneveau, D.~K. Lilly, Generalized smagorinsky model for
  anisotropic grids, Physics of Fluids A: Fluid Dynamics 5~(9) (1993)
  2306--2308.

\bibitem{spalding1961single}
A single formula for the law of the wall, Journal of Applied Mechanics 28~(3)
  (1961) 455--458.

\bibitem{bae2021effect}
H.~J. Bae, A.~Lozano-Dur{\'a}n, Effect of wall boundary conditions on a
  wall-modeled large-eddy simulation in a finite-difference framework, Fluids
  6~(3) (2021) 112.

\bibitem{liefvendahl2017formulation}
M.~Liefvendahl, T.~Mukha, S.~Rezaeiravesh, Formulation of a wall model for LES
  in a collocated finite-volume framework, Department of Information
  Technology, Uppsala University, 2017.

\bibitem{larsson2016large}
J.~Larsson, S.~Kawai, J.~Bodart, I.~Bermejo-Moreno, Large eddy simulation with
  modeled wall-stress: recent progress and future directions, Mechanical
  Engineering Reviews 3~(1) (2016) 15--00418.

\bibitem{yang2017log}
X.~I. Yang, G.~I. Park, P.~Moin, Log-layer mismatch and modeling of the
  fluctuating wall stress in wall-modeled large-eddy simulations, Physical
  review fluids 2~(10) (2017) 104601.

\bibitem{lee2023artificial}
Y.~M. Lee, J.~H. Lee, J.~Lee, Artificial neural network-based wall-modeled
  large-eddy simulations of turbulent channel and separated boundary layer
  flows, Aerospace Science and Technology 132 (2023) 108014.

\bibitem{zhuang2021learned}
J.~Zhuang, D.~Kochkov, Y.~Bar-Sinai, M.~P. Brenner, S.~Hoyer, Learned
  discretizations for passive scalar advection in a two-dimensional turbulent
  flow, Physical Review Fluids 6~(6) (2021) 064605.

\end{thebibliography}

\end{document}